\documentclass[fleqn,10pt]{wlscirep}
\usepackage[utf8]{inputenc}
\usepackage[T1]{fontenc}
\usepackage{placeins} 
\usepackage{commath}

\title{Accurate  Clinical Toxicity Prediction using Multi-task Deep Neural Nets and Contrastive Molecular Explanations}

\author[1]{Bhanushee Sharma}
\author[2]{Vijil Chenthamarakshan}
\author[2]{Amit Dhurandhar}
\author[3]{Shiranee Pereira}
\author[4]{James A. Hendler}
\author[1, *]{Jonathan S. Dordick}
\author[2, *]{Payel Das}

\affil[1]{Chemical and Biological Engineering, RPI, Troy, NY, USA}
\affil[2]{IBM Research, Yorktown Heights, NY, USA}
\affil[3]{People For Animals, Chennai}
\affil[4]{Computer Science, RPI, Troy, NY, USA}

\affil[*]{Corresponding: daspa@us.ibm.com, dordick@rpi.edu}

\begin{abstract}
    Explainable machine learning for molecular toxicity prediction is a promising approach for efficient drug development and chemical safety. A predictive ML model of toxicity can reduce experimental cost and time while mitigating ethical concerns by significantly reducing animal and clinical testing. Herein, we use a deep learning framework for simultaneously modeling \textit{in vitro}, \textit{in vivo}, and clinical toxicity data. Two different molecular input representations are used; Morgan fingerprints and pre-trained SMILES embeddings.  A multi-task deep learning model accurately predicts toxicity for all endpoints, including clinical, as indicated by the area under the Receiver Operator Characteristic curve and balanced accuracy. In particular, pre-trained molecular SMILES embeddings as input to the multi-task model improved clinical toxicity predictions compared to existing models in  MoleculeNet benchmark. Additionally, our multitask approach is comprehensive in the sense that it is comparable to state-of-the-art approaches for specific endpoints in \textit{in vitro}, \textit{in vivo} and clinical platforms. Through both the multi-task model and transfer learning, we were able to indicate the minimal need of \textit{in vivo} data for clinical toxicity predictions. To provide confidence and explain the model's predictions, we adapt a post-hoc contrastive explanation method that returns pertinent positive and negative features, which correspond well to known mutagenic and reactive toxicophores, such as unsubstituted bonded heteroatoms, aromatic amines, and Michael receptors. Furthermore,  toxicophore recovery by pertinent feature analysis captures more of the \textit{in vitro} (53\%) and \textit{in vivo} (56\%), rather than of the clinical (8\%), endpoints, and indeed uncovers a preference in known toxicophore data towards \textit{in vitro} and \textit{in vivo} experimental data. To our knowledge, this is the first contrastive explanation, using both \textit{present} and \textit{absent} substructures, for predictions of clinical and \textit{in vivo} molecular toxicity. 
    
\end{abstract}

\begin{document}

\flushbottom
\maketitle

\thispagestyle{empty}

    Toxicity remains a major driver of drug candidate failure in drug development, resulting in the high cost of drugs that make it into the market \cite{Hwang2016, Hay2014}. This phenomenon has persisted despite the surge in new chemical entities (NCEs) resulting from both advances in omics technology and the ability of Machine Learning (ML) models to generate novel molecules \cite{chenthamarakshan2020cogmol, Zhavoronkov2019, Li2018}. Consequently, there is an increasing need to accurately and efficiently predict the safety of new drug candidates in humans. To this end, there has been an escalation of ML models predicting toxicity, but not without its challenges. Of the many challenges, one is to correctly model a multi-faceted problem across different \textit{in vitro}, \textit{in vivo} and clinical platforms of varying granularities. Another non-trivial challenge is to comprehensively explain predictions across all these platforms which is non-trivial. To illustrate on the first challenge, a variety of ML models have been applied to chemical, biological, and mechanistic data that predict the toxicity of chemicals with varying granularity and relevancy to toxicity in humans, i.e., clinical toxicity (Section 1 in Supplementary). These models have differed in inputs \cite{Luco1997,Abdelaziz2016, Mayr2016, Matsuzaka2019, Fernandez2018}, architectures \cite{Ajmani2006,Chavan2015, Cao2012, Polishchuk2009, Mayr2016,Jimenez-Carretero2018}, and prediction platforms (i.e., endpoints or the specific experimental target including \textit{in vivo}, \textit{in vitro}, or clinical). A majority of ML models have focused on predicting specific \textit{in vitro} endpoints \cite{Huang2016, tox21,Tice2013,Kavlock2009} from only chemical structures , differing mainly in the molecular representations used \cite{Mayr2016, Matsuzaka2019,Fernandez2018, Altae-Tran2017}. In particular, nuclear receptor endpoints were a major focus of the Tox21 Challenge, a data challenge as a subset of the broader “Toxicology in the 21st Century” initiative. The Tox21 challenge provides the results of 12 \textit{in vitro} assays that test seven different nuclear receptor signaling effects and five stress response effects of 10,000 molecules in cells \cite{Huang2016, tox21,Tice2013,Kavlock2009}.
    
    Toxicities predicted \textit{in vitro} or \textit{in vivo} are not necessarily in concordance with each other \cite{Cox2016, Otava2015} nor to humans \cite{Hay2014, Olson2000, Martin2012, Tamaki2013, Becker2017}, thus reducing their ability to predict clinical toxicity \cite{Hay2014}. The granularity of toxicity tests varies across the \textit{in vitro}, \textit{in vivo}, and clinical platforms. \textit{In vitro} testing is the most granular and captures the ability of a chemical to disrupt biological pathways at the cellular level. In contrast,clinical testing is coarse-grained and captures the interactions of chemicals at multiple levels in the human, including organs and tissues. Thus, ML models trained on \textit{in vitro} and \textit{in vivo} data might not reliably capture clinical toxicity.
    
    Despite toxicity being a multi-task problem, majority of ML models have predicted toxicity in each platform separately with single-task models (Section 1 in Supplementary). A single molecule can demonstrate simultaneously a multitude of responses in different assays and different living organisms. Various solutions for modeling multiple toxic endpoints have been reported, by creating separate binary classification models for each endpoint \cite{Liua, Abdelaziz2016}, or by using multiple classification models the define classes \cite{Gadaleta2019, Li2014, Idakwo2019, Chen2013,Jiang2015, Raies2018, Wu2018, Sosnin2019, Xu2017}. Yet, thus far, the multiplicity problem has been modeled by predicting multiple endpoints within the same testing platform: \textit{in vitro}, \textit{in vivo}, or in humans, separately.
    
    The second challenge of ML models in predictive toxicology is to explain the predictions made, particularly within deep learning-based models. ML toxicity models have increasingly shifted towards deep learning \cite{Mayr2016,Jimenez-Carretero2018} pushed by its superior predictive performance \cite{Mayr2016} and ability to self-select significant features \cite{Tang2018}. Deep learning (DL) models are ``black-box'' models with limited explainability, i.e., do not provide reasoning for predicting that a molecule is toxic or nontoxic. This explanation is essential for designing new molecules and for providing greater confidence to the experimentalist end-users. As a result, the Organisation for Economic Co-operation and Development (OECD) strongly recommends that predictions of computational toxicology models be explainable \cite{OECD_guidance}. Efforts in explaining toxicity predictions have focused on pinpointing the \textit{presence} of certain features, such as toxicophores, derived from a range of methods, including simpler quantitative structure-activity relationship (QSAR) models \cite{Raies2016, Sharma2017} and explaining training in Deep Neural Networks (DNNs)\cite{Mayr2016} (Section 1.1 in Supplementary). These methods generally do not examine the effect of the \textit{absence} of these features. Defining minimal and necessary features, as well as present \textit{and} absent features, might provide a  comprehensive explanation that is more intuitive to  end-users. 
    
    Herein, we have developed a deep learning framework with the aim of improving accuracy and explainability of clinical toxicity predictions by taking advantage of \textit{in vitro}, \textit{in vivo}, and clinical toxicity data, and more advanced molecular representations. We simultaneously predicted \textit{in vitro}, \textit{in vivo}, and clinical toxicity through deep multi-task models, while comparing to their single-task and transfer learning counterparts. Two different molecular representations were tested; Morgan fingerprints and in-house created pre-trained SMILES embeddings encoding for the relationships among the chemicals. We used this framework for establishing concordance across \textit{in vitro}, \textit{in vivo}, and clinical datasets. Notably, in response to the adoption of the 3 Rs (Replacement, Reduction, and Refinement of animal testing) in global legislation \cite{Russell1959, Tornqvist2014}, we assessed the need for animal data in making clinical toxicity predictions. To provide more comprehensive molecular explanations of toxicity predictions, we adopted the Contrastive Explanations Method (CEM) explainability model \cite{Dhurandhar2018} which explains "black-box" DNN predictions by revealing pertinent positive (PP) and pertinent negative (PN) features as chemical structures correlating to a given prediction. Specifically, we explained single-task toxicity predictions \textit{in vitro}, \textit{in vivo}, and clinically. The PPs represent the minimum required substructures for classification of a molecule (toxicophores for a toxic prediction), and the PNs represent the minimum changes to a molecule that would flip its predicted class label, from toxic to nontoxic or vice versa. Such an explanation should expand the scope of predictive toxicology while providing information on both toxicophores and nontoxic substructures.

\section{Model Framework}

\begin{figure}[!htb]
    \vspace{-0.5em}
    \centering
    \makebox[0pt]{%
    \includegraphics[scale=0.55]{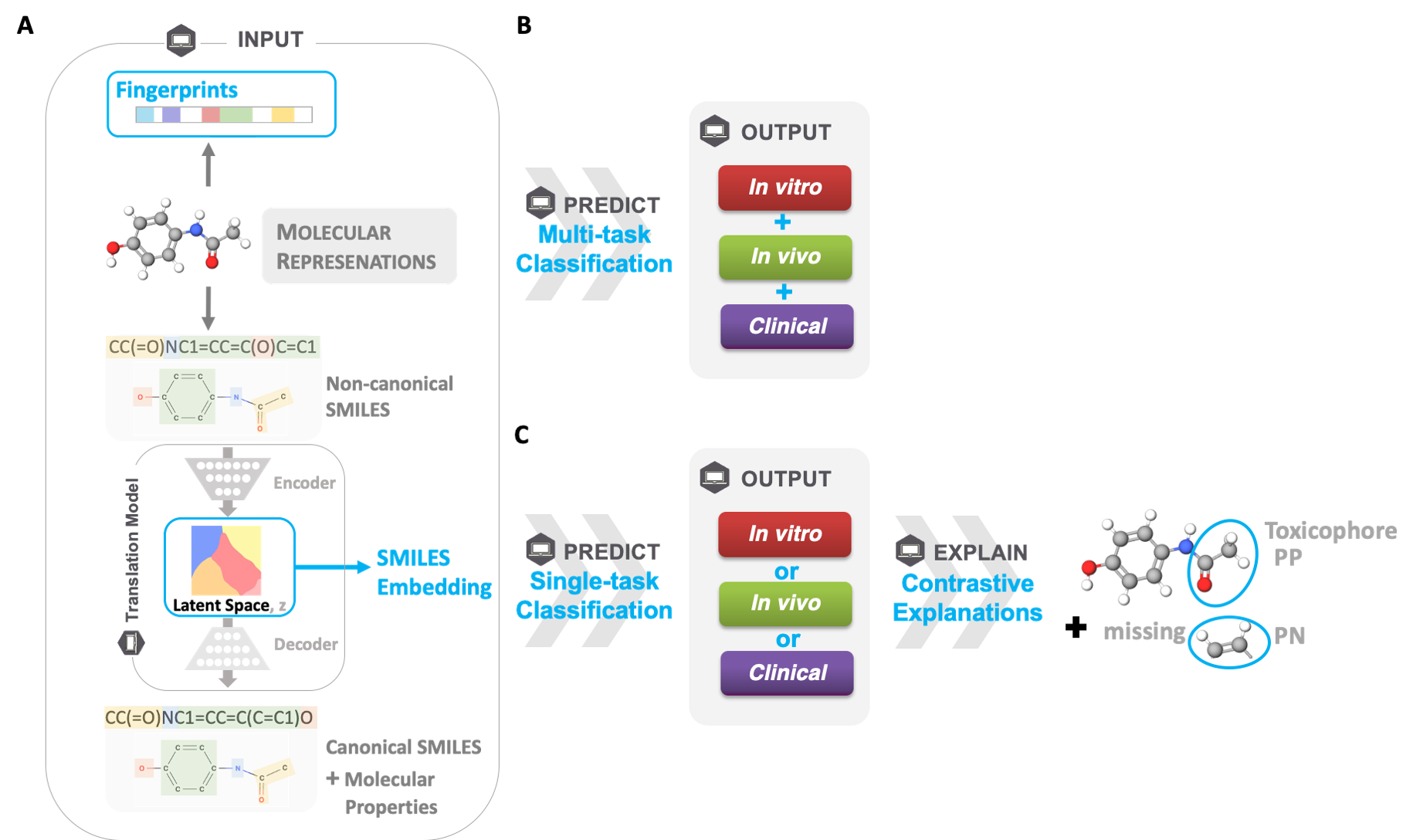}}
    \caption{\textbf{Framework adopted in this study for explainable, single-task, and multi-task prediction of \textit{in vitro}, \textit{in vivo}, and clinical toxicity.} A) Given an input of different molecular representations, fingerprints and latent space SMILES embeddings, B) a multi-task classification model predicts whether a molecule is toxic or not for \textit{in vitro}, \textit{in vivo}, and clinical endpoints. Furthermore, the Contrastive Explanation method explains C) predictions from single-task models trained on fingerprints for the same endpoints. The method pinpoints minimal and necessary chemical substructures that are either present (pertinent positive, PP) or absent (pertinent negative, PN) for a specific prediction.
}
    \label{fig:framework}
\end{figure}

Two different molecular representations were used as inputs; commonly used Morgan Fingerprints (FP) and more complex SMILES embeddings (SE) (Fig~\ref{fig:framework}(A)). FP vectorize the presence of a substructure within varying radii around an atom. FP are easy to compute and have high performance among other fingerprints \cite{OBoyle2016}, and are thus widely used. However, FP are simplistic representations of chemical structures, not coding for relationships between substructures, unlike molecular graphs, nor for relationships between the chemicals.  To improve on this, the SE were created using a neural network-based model that translates from non-canonical SMILES to canonical SMILES, encoding for the relationship between chemicals within the datasets. 

With these molecular representations, toxicities \textit{in vivo}, \textit{in vitro}, and clinical, were predicted through multi-task (MTDNN) (Fig~\ref{fig:framework}(B)) and single-task Deep Neural Networks (STDNN) (Fig~\ref{fig:framework}(B)). The MTDNN predicts each platform (\textit{in vivo}, \textit{in vitro}, clinical) as a different task within one model, with each task consisting of a single or multiple classes (or endpoints). In contrast, the STDNN predicts each platform with a separate model. 

As a proof-of-concept, endpoints from previous benchmarking efforts \cite{Wu2017a} and data challenges \cite{Huang2016} were chosen, and are not exhaustive. For the clinical platform, the endpoint was whether or not a molecule failed clinical phase trials due to toxicity, as obtained from the ClinTox dataset \cite{Wu2017a}. For the \textit{in vitro} platform, 12 different endpoints from the Tox21 Challenge\cite{Huang2016} were used, whether or not a molecule is active in disrupting seven neural receptor assays and five stress response assays. Finally, for the \textit{in vivo} platform, one endpoint for acute oral toxicity in mice was used from the commercially available RTECS (Registry of Toxic Effects of Chemical Substances) dataset. The acute oral toxicity endpoint was defined by an LD\textsubscript{50} (lethal dose for 50\% of the population) cutoff of 5,000 mg/kg specified by GHS and EPA ($<$ 5,000 mg/kg as toxic,  $>$ 5,000 mg/kg as nontoxic).

We further tested the need of \textit{in vivo} data to predict clinical toxicity with the multi-task DNN and its transfer learning counterpart. Different combinations of \textit{in vivo}, \textit{in vitro}, and clinical tasks in the MTDNN were investigated to determine the most relevant tasks for predicting clinical toxicity. One way to achieve this is to leverage transfer learning, which allows a base model trained on one task to be re-purposed for another related task. In our use case, we compared the ability of a base model trained on \textit{in vivo} or \textit{in vitro} data to be transferred to predicting clinical toxicity.

Further, we explained DNN predictions by adapting the Contrastive Explanations Method (CEM) \cite{Dhurandhar2018} to molecular structure input. Specifically, the CEM was adapted to explain toxicity predictions made by the STDNN  trained on Morgan fingerprints for \textit{in vivo}, \textit{in vitro}, and clinical platforms (Fig~\ref{fig:framework}(C)). For an overview of related work on explanations for molecular toxicity prediction see Section 1.1 in Supplementary. We had previously performed a proof-of-concept \cite{Lim2020} for this approach by explaining predictions on one specific \textit{in vitro} endpoint ("SR-MMP", the ability of molecules to disrupt the mitochondrial membrane potential (MMP) in cells \cite{Huang2016}). In this work, we further expand to other \textit{in vitro} endpoints, and to \textit{in vivo} and clinical platforms. The CEM explains by identifying pertinent positive (PP) and counterfactual pertinent negative (PN) substructures within the input molecules. The PPs are the minimal and necessary substructures that correlate to a prediction, while the PNs are substructures that would switch the given prediction. 

\FloatBarrier

\section{Results}

\subsection{Deep Single-Task and Multi-Task Predictive Models - Morgan Fingerprints and SMILES Embeddings} \label{mtdnn}

We evaluated the performance of our framework by metrics of AUC-ROC (Area under Receiver Operating Characteristic curve) and balanced accuracy. We compared the performance of STDNN (blue in Figs~\ref{fig:auc-roc-moleculenet-embed}, \ref{fig:test-ba}) to MTDNN (multiple colors in Figs~\ref{fig:auc-roc-moleculenet-embed}, \ref{fig:test-ba}) with either SMILES embeddings (SE, darker color in Figs~\ref{fig:auc-roc-moleculenet-embed}, \ref{fig:test-ba}) or Morgan fingerprints (FP, lighter colors in Figs~\ref{fig:auc-roc-moleculenet-embed}, \ref{fig:test-ba}) as input. Performance of different platform combinations in the MTDNN was contrasted by combining all three \textit{in vivo}, \textit{in vitro} and clinical platforms (red in Figs~\ref{fig:auc-roc-moleculenet-embed}, \ref{fig:test-ba}), clinical and \textit{in vitro} platforms (purple in Fig~\ref{fig:auc-roc-moleculenet-embed}, \ref{fig:test-ba}), clinical and \textit{in vivo} platforms (orange in Figs~\ref{fig:auc-roc-moleculenet-embed}, \ref{fig:test-ba}), and \textit{in vitro} and \textit{in vivo} platforms (yellow in Figs~\ref{fig:auc-roc-moleculenet-embed}, \ref{fig:test-ba}). 

Area under the ROC curves (AUC-ROC) was determined against the best performing models in MoleculeNet \cite{Wu2018} (grey in Fig~\ref{fig:auc-roc-moleculenet-embed}), a widely used benchmark for molecular predictions. The best model employed by MoleculeNet on ClinTox (clinical) was a graph neural net baseline operating on molecular graphs \cite{weave} (Weave). On Tox21 (\textit{in vitro)}, the best performing model in Moleculenet was a graph convolutional neural net (GC)\cite{Wu2018}. MoleculeNet did not benchmark RTECS (\textit{in vivo)}, but we used their provided benchmark methods to train and test the models used on ClinTox and Tox21\cite{Wu2018}, on RTECS. The resulting best model from MoleculeNet on RTECS was the influence relevance voting system (IRV), which is an enhanced k-nearest neighbor model augmented by weights provided from one-layer DNNs\cite{Wu2018}. 

\subsubsection{Area under Receiver Operating Characteristic curve} \label{auc}
\begin{figure}[!htb]
    \vspace{-0.5em}
    \centering
    \makebox[0pt]{%
    \includegraphics[scale=0.45]{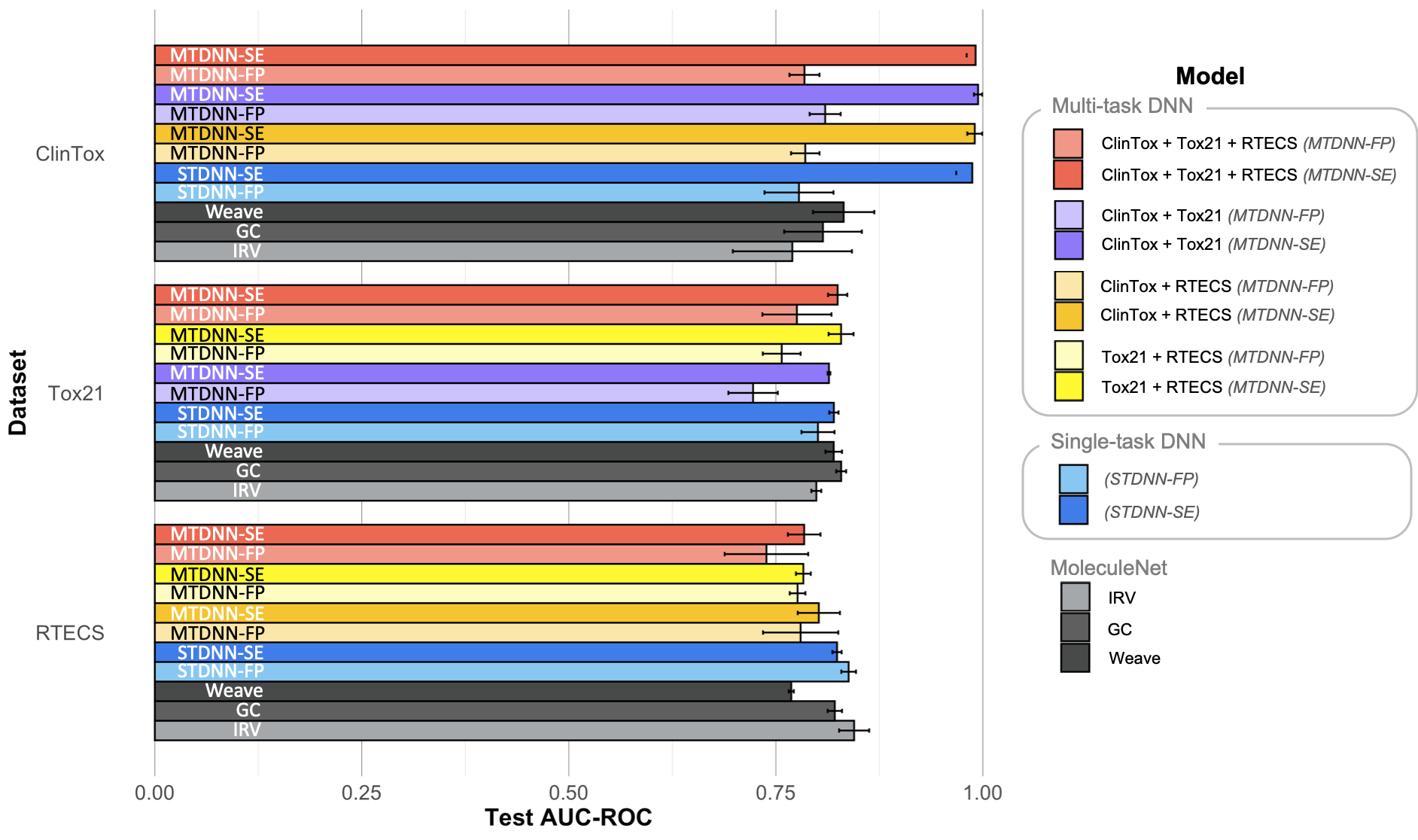}}
    \caption{\textbf{Test AUC-ROC values for ClinTox, Tox21, and RTECS predictions, comparing multi-task models to single-task and baseline MoleculeNet models, with SMILES embeddings and Morgan fingerprints as inputs.} The best performing model on ClinTox, Tox21, and RTECS from MoleculeNet is displayed. All other MoleculeNet models are in Supplementary Fig S4. 
}
    \label{fig:auc-roc-moleculenet-embed}
\end{figure}

AUC-ROC performance markedly improved on predicting clinical toxicity (ClinTox) using the combination of SE and MTDNN. Compared to Weave, the best performing baseline model on ClinTox (dark gray in Fig~\ref{fig:auc-roc-moleculenet-embed}), the single-task DNN with SE (STDNN-SE, dark blue in Fig~\ref{fig:auc-roc-moleculenet-embed}) improved AUC-ROC values from 0.832 $\pm$ 0.037 to 0.987 $\pm$ 0.019. The multi-task DNN trained with SE (MTDNN-SE) further improved AUROC performance on ClinTox with similar values when trained on all three platforms (0.991 $\pm$ 0.011, dark red in Fig~\ref{fig:auc-roc-moleculenet-embed}) or trained on ClinTox and Tox21 (0.994 $\pm$ 0.005, dark purple in Fig~\ref{fig:auc-roc-moleculenet-embed}). For predicting \textit{in vitro} toxicity (Tox21), the MTDNN and STDNN models performed similarly to MoleculeNet's GC model (0.829 $\pm$ 0.006, grey in Fig~\ref{fig:auc-roc-moleculenet-embed}) when using SE as input. STDNN-SE gave AUC-ROC of 0.820 $\pm$ 0.006 (dark blue in Fig~\ref{fig:auc-roc-moleculenet-embed}). The MTDNN-SE performed comparably trained on all three platforms (0.825 $\pm$ 0.012, dark red in Fig~\ref{fig:auc-roc-moleculenet-embed}) and on Tox21 and RTECS (0.829 $\pm$ 0.015, dark yellow in Fig~\ref{fig:auc-roc-moleculenet-embed}). Interestingly, for the \textit{in vivo} endpoint (RTECS), neither the multi-task model nor the use of the SE showed improved  performance when compared to the fingerprint-based single-task models (light blue in Fig~\ref{fig:auc-roc-moleculenet-embed}) or MoleculeNet's IRV model (light grey in Fig~\ref{fig:auc-roc-moleculenet-embed}). Thus, the MTDNN-SE appeared to improve clinical toxicity predictions and provided comparable performance on Tox21, but not on RTECS.
 
\FloatBarrier

\subsubsection{Balanced Accuracy}\label{ba}

Skewed datasets are a prevalent problem in predictive toxicology \cite{Sedykh2011, Thomas2012, Abdelaziz2016}. Regardless of the platform, the distribution of toxic and nontoxic examples is often imbalanced. Within the datasets studied here, the imbalance is biased towards the "nontoxic" class in ClinTox and Tox21, and the "toxic" class in RTECS (Fig S5 in Supplementary for single-task models, and Fig S6 for multi-task models). This biases the AUC-ROC values towards a small fraction of true toxic or true nontoxic predictions. The balanced accuracy (BA) mertic takes into account this imbalance and has been used as a more representative metric for predictive toxicology models \cite{Sedykh2011, Thomas2012, Abdelaziz2016}. Balanced accuracy averages the sensitivity and the specificity. The former is the fraction of correctly classified positive classes out of all possible positives in the dataset, i.e., fraction of true positives out of correctly classified positives and falsely classified negatives. Conversely, the specificity is this measure for the true negatives of the model, i.e., the fraction of true negatives correctly classified out of the total number of negatives in the dataset (both the correctly classified negatives and the falsely classified positives). However, current baseline models in MoleculeNet  do not provide balanced accuracy performance on the toxicity benchmarks.

\begin{figure}[!htb]
    \vspace{-0.5em}
    \centering
    \makebox[0pt]{%
    \includegraphics[scale=0.5]{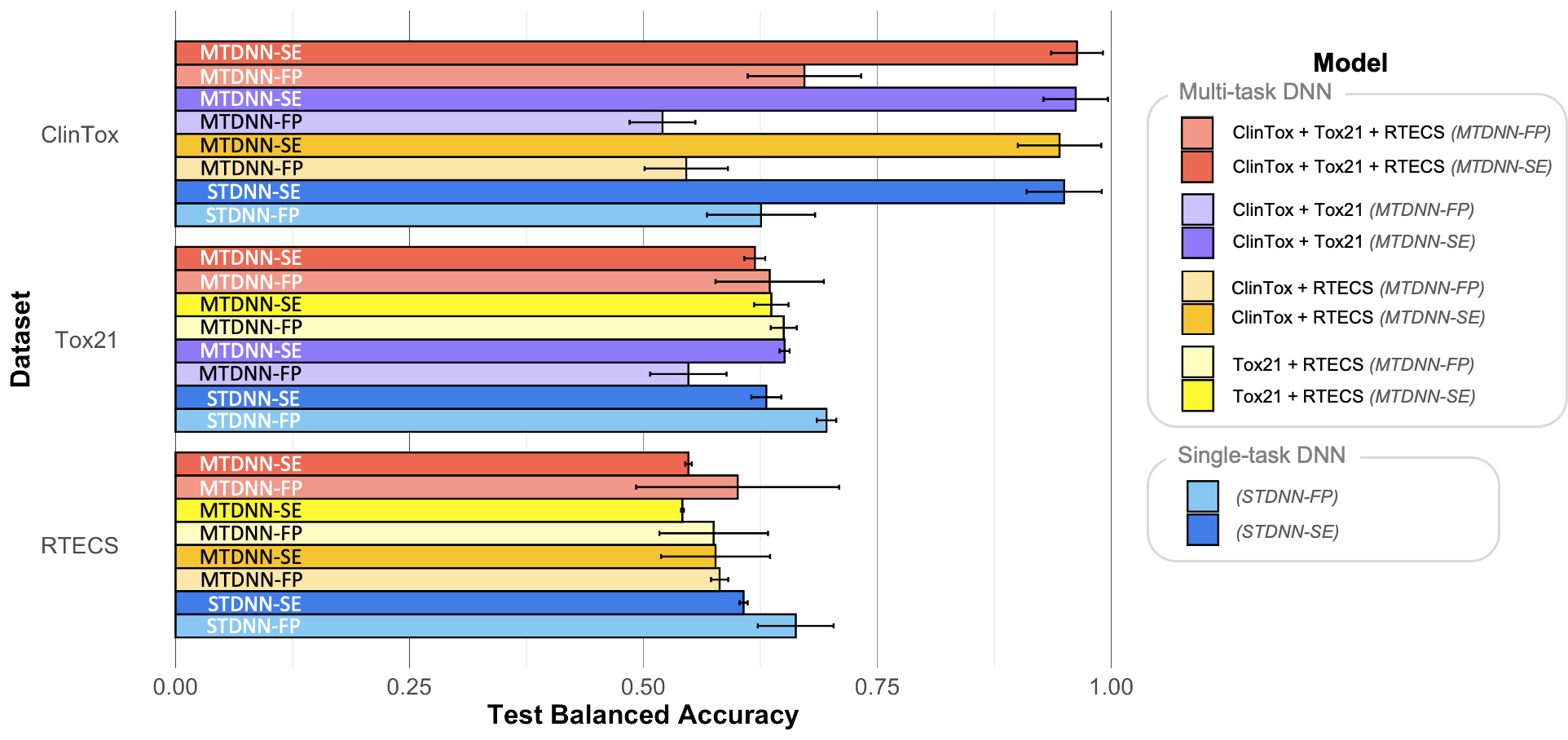}}
    \caption{\textbf{Average Balanced Accuracy on the test set for ClinTox, Tox21 and, RTECS predictions, comparing multi-task models to single-task models, with SMILES embeddings and Morgan fingerprints as inputs.}}
    \label{fig:test-ba}
\end{figure}

We report balanced accuracy for all three platforms (Fig~\ref{fig:test-ba}). High balanced accuracy, 0.95-0.96, was achieved by both MTDNN-SE and STDNN-SE models for clinical toxicity predictions (ClinTox, dark colors in Fig~\ref{fig:test-ba}). The best balanced accuracy on the clinical task was given by the MTDNN-SE trained on all three platforms (0.963 $\pm$ 0.028, dark red in Fig~\ref{fig:test-ba}).  The DNN models also resulted in a balanced accuracy of $\sim$ 0.64  on Tox21 and RTECS predictions, which is  still notably better than random classification (balanced accuracy of 0.5) (Fig~\ref{fig:test-ba}). Overall, use of both the multi-task setting and SMILES embeddings helped improve the balanced accuracy on the clinical platform much more so than on \textit{in vitro} or \textit{in vivo}.

\FloatBarrier

\subsection{Testing the relative importance of \textit{in vivo} data to predict clinical toxicity} \label{invivo}

The relative importance for \textit{in vivo} data for predictions of clinical toxicity was assessed by training different combinations of platforms in the multi-task model, with or without \textit{in vivo} data, and its transfer learning counterpart. For clinical toxicity predictions (ClinTox), MTDNN models without \textit{in vivo} data (purple in Fig~\ref{fig:auc-roc-moleculenet-embed}) performed better in terms of AUC-ROC values when compared to models with \textit{in vivo} data (red or orange in  Fig~\ref{fig:auc-roc-moleculenet-embed}). To further examine this correlation, the distribution of true/false positives and true/false negatives of clinical toxicity predictions was determined using true labels based on \textit{in vitro} or \textit{in vivo} datasets (Fig S11). Conclusions on this correlation are difficult due to the small number of chemicals common across the platforms. From the small overlap of chemicals, a larger number of true positives than false positives was determined for the predictions on ClinTox using true labels from Tox21 than using RTECS (Fig S11). The SR-p53 \textit{in vitro} assay (stress response on p53) in Tox21 was the sole exception. 

\begin{figure}[!htb]
    \vspace{-0.5em}
    \centering
    \makebox[0pt]{%
    \includegraphics[scale=0.7]{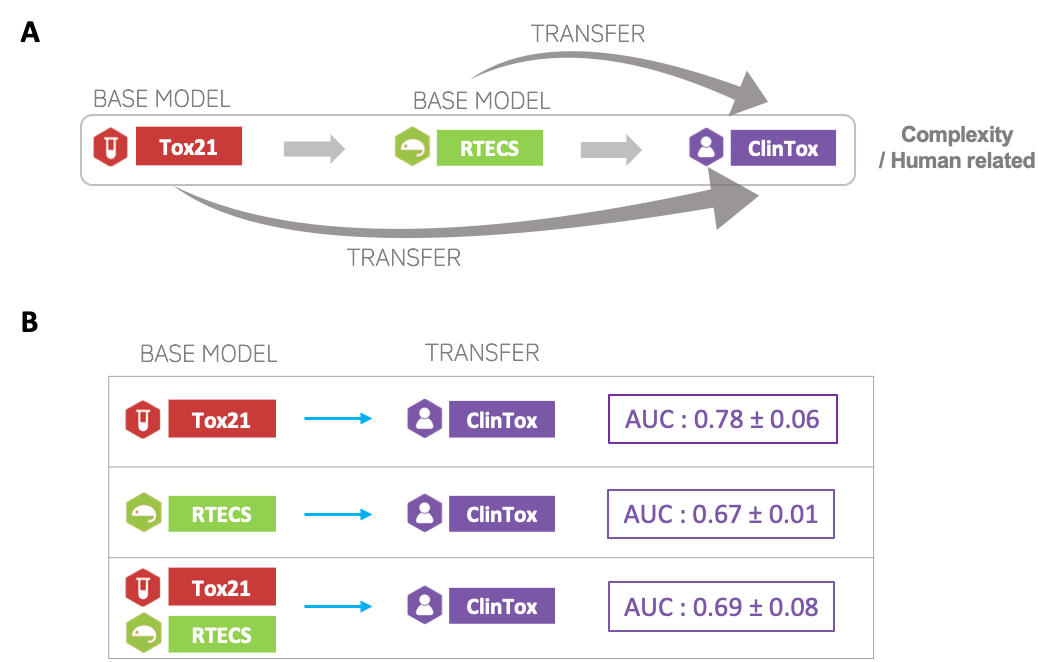}}
    \caption{\textbf{Comparing transfer learning to predicting clinical toxicity from base \textit{in vitro} and \textit{in vivo} models.} A) Schematic of transfer learning model, with a base model pre-trained on either on \textit{in vitro} or \textit{in vivo} data transferred to predicting clinical toxicity. B) AUC results on the ClinTox (clinical) tasks with one epoch training during transfer learning to ClinTox tasks.
}
    \label{fig:transfer}
\end{figure}

Transfer learning applies knowledge gained from pre-training on a base model to predicting in a related domain \cite{Weiss2016}. We contrasted the application of base models trained on \textit{in vivo} or \textit{in vitro} data to predicting clinical toxicity (Fig~\ref{fig:transfer}(A)). A base model trained with \textit{in vivo} (RTECS) data decreased in AUC-ROC performance on ClinTox (clinical) when compared to a base model containing only \textit{in vitro} (Tox21) data (AUC of 0.78 ± 0.06 versus 0.67 ± 0.01 or 0.69 ± 0.08) (Fig~\ref{fig:transfer}(B)). Thus, pre-training with an \textit{in vitro} base model transfers to predicting clinical toxicity better than with an \textit{in vivo} base model.

\begin{figure}[!htb]
    \vspace{-0.5em}
    \centering
    \makebox[0pt]{%
    \includegraphics[scale=0.5]{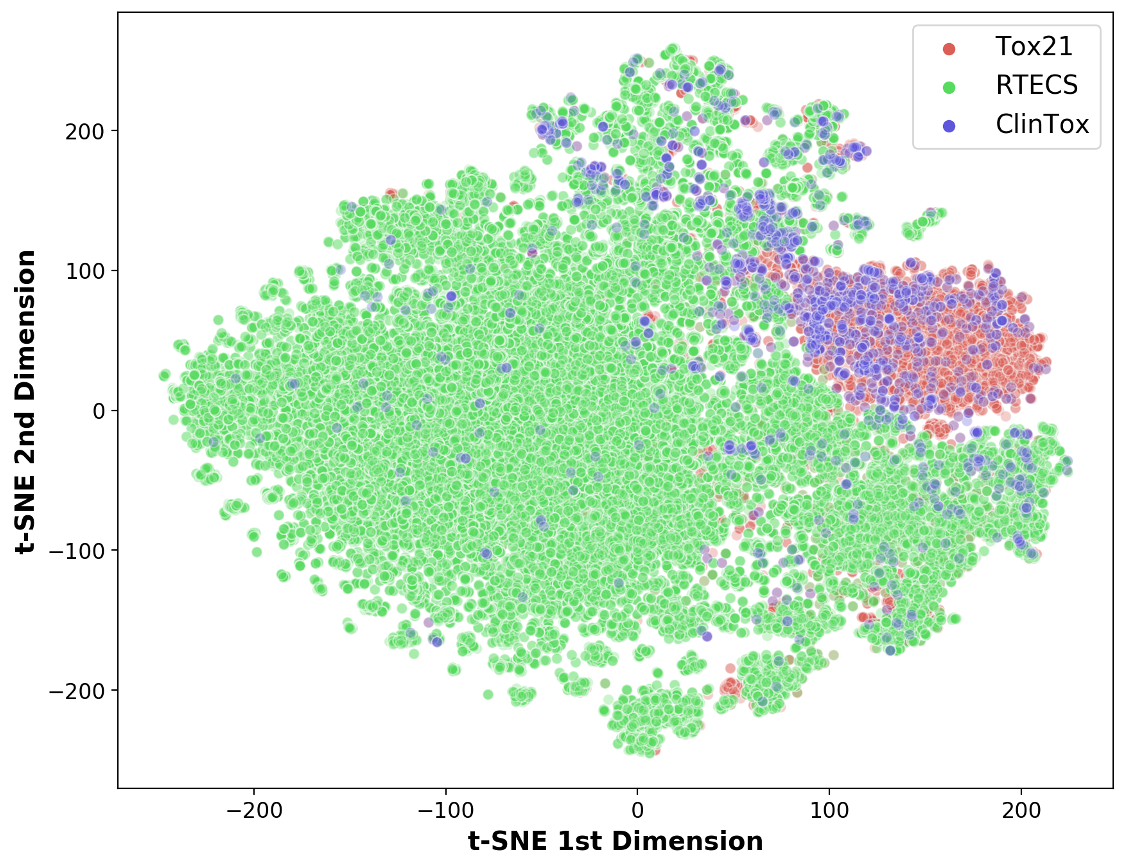}}
    \caption{\textbf{t-SNE of SMILES embeddings of chemicals in the Tox21, RTECS, and ClinTox datasets.} Distances correlate to similarities of the chemicals across these datasets; shorter the linear distance, the more similar are the chemicals. ClinTox chemicals overlap more with Tox21 chemicals than with RTECS. 
}
    \label{fig:tsne}
\end{figure}

To investigate whether the type of chemicals within the \textit{in vitro}, \textit{in vivo}, and clinical datasets affected the ability of \textit{in vivo} data to predict clinical toxicity, we visualized the relationships between the chemicals using t-distributed stochastic neighbor embeddings (t-SNE)\cite{Hinton}. t-SNE, a method that maps high-dimensional data to lower dimensions while preserving local similarities (i.e., distances between datapoints) \cite{Hinton}. We applied \textit{t}-SNE mapping to SE of the chemicals in the Tox21, RTECS, and ClinTox datasets, with each dot representing a chemical and distance representing similarity (Fig~\ref{fig:tsne}). The map is dominated by RTECS chemicals (green) due to the larger number of chemicals present in RTECS than both the Tox21 and ClinTox datasets. However, when examining overlap of the chemicals, the majority of the overlap is among ClinTox (purple) and Tox21 (red) chemicals, with some overlap between ClinTox (purple) and RTECS (green) chemicals. Thus, chemicals present in the clinical dataset (ClinTox) are more related to the chemicals present in the \textit{in vitro} dataset (Tox21) rather than those in the \textit{in vivo} (RTECS) dataset. 

\FloatBarrier

\FloatBarrier

\subsection{Contrastive Molecular-Level Explanations of Toxicity}\label{cem}

STDNN and MTDNN have improved the accuracy in predicting clinical toxicity. However, these DNNs cannot explain why a molecule was predicted to be toxic. To improve the trustworthiness of our results and to expand the current scope of explaining toxicity predictions, we adapted the contrastive explanation method (CEM) \cite{Dhurandhar2018} for molecular-level explanations of toxicity predictions. Specifically, we have adapted the CEM to explain \textit{in vitro}, \textit{in vivo} and clinical toxicity predictions, made by the STDNN trained on Morgan fingerprints (STDNN-FPs). STDNN-FPs were chosen as they can provide easy-to-understand substructure level explanations from their FP input, while maintaining consistent and significant accuracy across all platforms. 

The CEM more comprehensively explains DNN predictions by identifying \textit{present} (pertinent positive, PP) and \textit{absent} (pertinent negative, PN) substructures within the molecules that correlate to a prediction. For instance, for a molecule predicted to be "toxic", the PP substructures are the minimum and necessary substructures within the molecule that correlate to the "toxic" prediction. Conversely, the PN substructures represent the minimum and necessary substructures missing from the molecule that when added convert the "toxic" prediction to "nontoxic". To this end, we obtained PP and PN substructures for all of the molecules in the ClinTox, Tox21 and RTECS test sets. These PP and PN substructures illuminate the decisions made by the STDNN-FPs and provide substructures that correlate to "toxic" and "nontoxic" predictions. We focused on only correct "toxic" and "nontoxic" predictions in the test set of ClinTox, Tox21, and RTECS. The CEM collected PP and PNs for 277, 1545, 7602 correctly predicted chemicals in ClinTox, Tox21, and RTECS, respectively. For each molecule, ten PP and ten PN substructures were collected, totalling 94,240 PP and PN substructures. 

\subsubsection{Most common PP and PN Substructures for "toxic" and "nontoxic" molecules} \label{top10}

To focus on prevalent molecular features for explanations from the $\sim$ 94,000 substructures obtained, we analyzed the ten most common PP and PN substructures by frequency for correct "toxic" and "nontoxic" \textit{in vitro}, \textit{in vivo} and clinical predictions (top five are shown in Fig~\ref{fig:toxic}, top ten in Supplementary Fig S7+S8). These top ten most common PP and PNs represent the ten most common explanations for "toxic" and "nontoxic" predictions. However, for the highly skewed ClinTox test set, with only 1-2 "toxic" molecules, only the "nontoxic" predictions were examined. Along with the substructures, the CEM also provides the most significant (central) atom of the substructure (blue circles in Fig~\ref{fig:toxic}). 

Two approaches were used to obtain explanations for "toxic" predictions of molecules by the STDNN: (1) PP substructures present in "toxic" molecules, and (2) PN substructures missing from "toxic" molecules which, if present, would change the prediction from "toxic" to "nontoxic"  (Fig~\ref{fig:toxic} and Fig S7). From the top ten most frequent PPs of molecules predicted to be "toxic",  common substructures were identified for both Tox21 and RTECS, heavily involving nucleophilic N, O, and aryl groups (Fig~\ref{fig:toxic} and Fig S7). The most common was just a carbon fragment, perhaps due to the abundance of this substructure in the Morgan fingerprints. Substructures with oxygen were also common, either as a carbonyl (Tox21 and RTECS) or as an ether (RTECS). Different portions of aromatic rings were identified multiple times, within an aryl ring (Tox21 and RTECS), or within a benzyl group (Tox21 and RTECS). Finally, N was obtained for both Tox21 and RTECS.  Thus, the presence of nucleophilic N, O, and aryl containing substructures was commonly identified as explanations for \textit{in vitro} and \textit{in vivo} predictions of toxicity.

Explanations of "toxic" predictions obtained from the absence of PN substructures from molecules predicted to be toxic, most commonly contained carbon fragments and substructures with N and O for both Tox21 and RTECS, but also identified aryl halides for Tox21 and sulfur (S) containing substructures for RTECS (Fig~\ref{fig:toxic} and Fig S7). For O containing substructures, aromatic carboxylic (Tox21 and RTECS) and aliphatic carbxylic (Tox21) moieties were identified. N containing substructures were present as an aromatic amine (Tox21), as the heteroatom in cephalosporins (RTECS), an imine (RTECS), or a nitrite (Tox21). For substructures that differed between Tox21 and RTECS, Tox21 specified an aryl chloride, while RTECS specified S within an aromatic sulfonic acid group.

Similarly, the ten most common explanations for molecules predicted to be "nontoxic" were obtained (Fig~\ref{fig:toxic}, Fig S8). From the PP of molecules predicted to be "nontoxic", common substructures were identified for ClinTox, Tox21, and RTECS to be aryl fragments, and small substructures with N and O (Fig~\ref{fig:toxic}, Fig S8). Aryl fragments were present for Tox21 and ClinTox. Small substructures with O were frequent for all datasets, specifying only O, a hydroxyl, or a carbonyl. Substructures with only the N (Tox21, RTECS, ClinTox) or as an amide (RTECS) were also present. 

Explanations of "nontoxic" predictions obtained from the absence of PN substructures from molecules predicted to be nontoxic contained various complex aromatic substructures with N, O, S, P, I and F, and heavy metals (Tin, Sn and Mercury, Hg) (Fig~\ref{fig:toxic} and Fig S8). The difference among Tox21, RTECS, and ClinTox was in the complexity of the aryl structures and type of heteroatom. ClinTox had the most complex aryl substructures with different combinations of  N, O, P, or F, followed by RTECS with complex aryl substructures containing N, O, S, and the halide I. Finally, Tox21 had the least number of complex aromatic structures with N as a purine or an aromatic amide, with Hg as a substituent (organomercury), or with O as a phenol. N and O containing aliphatic substructures were also common through all the endpoints, while Tox21 also specified aliphatic substructures with Sn. 

In the process of explaining “toxic” and “nontoxic” predictions, the CEM obtains PP and PN substructures correlating to toxicity and nontoxicity, i.e. computationally extracting toxicophores and nontoxic substructures. PP of "toxic" predictions and PN of "nontoxic" predictions are both substructures that correlate to toxic predictions, i.e., toxicophores. Conversely, PP of "nontoxic" predictions and PN of "toxic" predictions are substructures correlating to nontoxic predictions, i.e., nontoxic substructures. Thus, the above identified PP and PN substructures are also the top ten most commonly identified toxicophores and nontoxic substructures by the CEM within each dataset. 

\begin{figure}[!htb]
    \vspace{-0.2em}
    \centering
    \makebox[0pt]{%
    \includegraphics[scale=0.73]{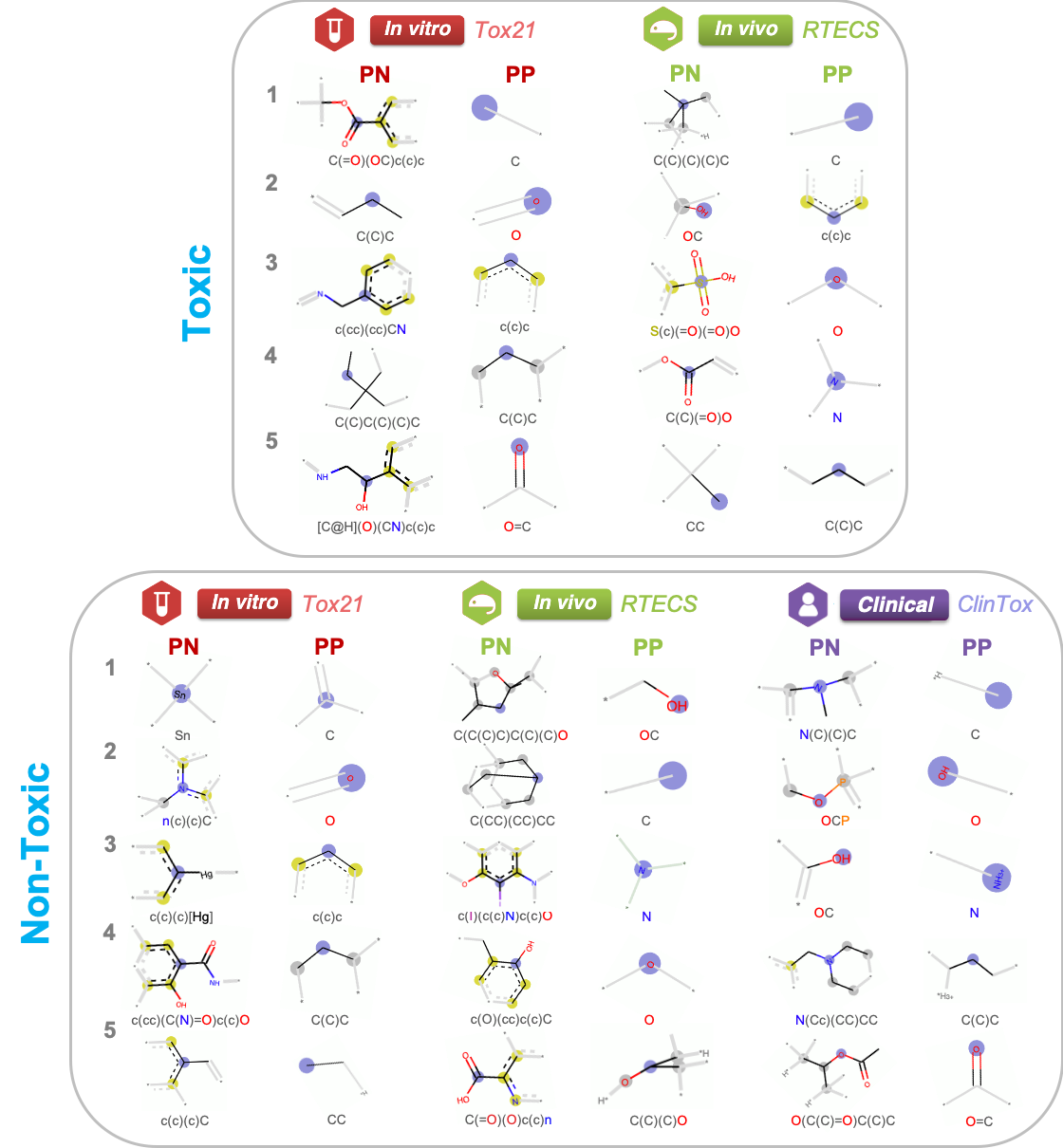}} 
    \caption{\textbf{Most common PP and PN substructures of correctly predicted toxic and nontoxic molecules across the Tox21, RTECS, and ClinTox endpoints.} ClinTox only had 1-2 examples of toxic molecules in the test set and thus was excluded here. All top 10 in Supplementary Fig S7 and S8.}
    \label{fig:toxic}
\end{figure}%
\FloatBarrier

\subsubsection{Verification of Pertinent features extracted to known toxicophores} \label{verification}

To verify the correlation between the obtained PP and PN substructures and the toxicity predictions, we matched the PP and PN substructures correlating to toxicity with known toxicophores. PP and PNs correlating to toxicity, or toxicophores, are obtained from the CEM by: (1) PPs of molecules correctly predicted to be "toxic", and (2) PNs of molecules correctly predicted to be "nontoxic" that would flip the molecule to be classified as "toxic". 

\begin{figure}[!htb]
    \vspace{-0.2em}
    \centering
    \makebox[0pt]{%
    \includegraphics[scale=0.6]{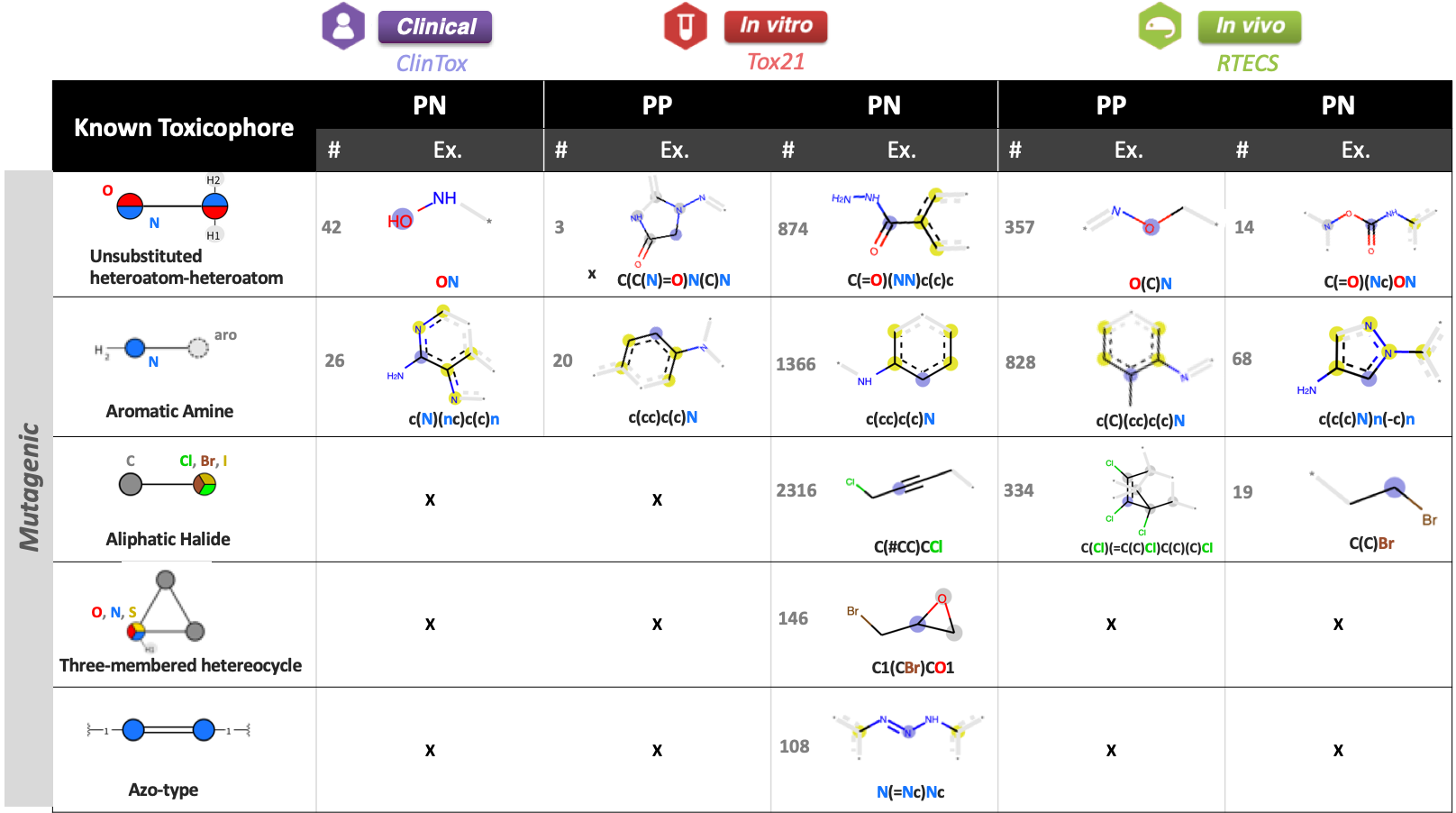}}
    \caption{\textbf{Matched Toxicophores.} Top three (ClinTox, RTECS) or top five (Tox21) matched known toxicophores to toxicophores collected from the CEM as PP of toxic molecules and PN of nontoxic molecules. For Tox21, the top five most frequent matches were examined due to the large number of matches. Three types of known toxicophores were matched: experimental and computational mutagenic toxicophores, and reactive substructures commonly used to filter molecules.  The table provides the number (\#) of matches, and specific examples (Ex.). Only mutagenic toxicophores are displayed in table, full list is given in Supplementary Fig S9.}
    \label{fig:matched}
\end{figure}

The literature contains a vast and diverse array of known toxicophores. Mutagenic toxicophores, in particular, have been widely used to verify results of computationally predicted toxicophores \cite{Mayr2016, Yang2017}. Here, we matched the toxicophores obtained from the CEM to known mutagenic toxicophores collected \textit{in vitro} experimentally \cite{Kazius2005} (Fig~\ref{fig:matched}), or computationally \cite{Yang2017} (Fig S9), and known reactive substructures commonly used to filter molecules \cite{Hevener2018} (Fig S9). The CEM was able to identify toxicophores across all these types of known toxicophores, both from PP substructures of correctly predicted toxic molecules and PN substructures of correctly predicted nontoxic molecules.

\FloatBarrier 

Examining the top three (ClinTox, RTECS) or the top five (Tox21) most frequently matched toxicophores, the CEM identified known toxicophores that are common to all endpoints (purple in Fig~\ref{fig:matched}), to ClinTox and Tox21 (magenta in Fig~\ref{fig:matched})), to Tox21 and RTECS (orange in Fig~\ref{fig:matched}), or unique to Tox21 (red in Fig~\ref{fig:matched})) or to RTECS (green in Fig~\ref{fig:matched})). For Tox21, top five most frequent matches were examined due to the large number of matches. Both Tox21 and RTECS recovered toxicophores common with ClinTox (unsubstituted heteroatom-heteroatoms, aromatic amines, and Michael receptors); however, only Tox21 identified thioesters, a reactive substructure present in ClinTox. Aromatic nitro, aliphatic halide, alkyl halide, heteroatom-heteroatom single-bond substructures were matched toxicophores found in both Tox21 and RTECS, but not in ClinTox. Uniquely, PPs and PNs in Tox21 matched to reactive cyanide, three-membered heterocycle, azo-type, carbonyl/ether, epoxides, thioepoxides, and disulfide substructures, while only RTECS matched to 1,2-dicarbonyls within the top three to five matched toxicophores. Notably, a larger number of toxicophore matches were found \textit{in vitro} (in the 1,000s) or \textit{in vivo} (in the 100s), compared to the clinical endpoint (in the 10s).

Thus far, we have matched known toxicophores to only PP and PNs correlating to toxicity. To discern whether the CEM correctly pinpoints PP and PNs substructures correlating to toxicity, we further matched all collected PP and PNs to known toxicophores. We expect to see a larger number of matches with PP and PN substructures correlating to toxicity (toxicophores), than to the converse (nontoxic substructures). Indeed, for ClinTox and Tox21, but not for RTECS, there are a larger number of matches to known toxicophores with CEM-derived toxicophores than with CEM-derived nontoxic substructures (Fig S10).

\section{Discussion}

We have demonstrated an improvement in predicting clinical toxicity using pre-trained SMILES embeddings as input molecular representations within a multi-task deep neural network that simultaneously learns \textit{in vitro}, \textit{in vivo}, and clinical toxicity tasks.  Through our multi-task model, we investigated the benefits of learning from more diverse toxicity data in \textit{in vitro}, \textit{in vivo}, and clinical platforms for predicting clinical toxicity. We also leveraged pre-trained molecular SMILES embeddings as inputs, which better captured intermolecular relationships within a larger corpus. Compared to the existing MoleculeNet baseline and our single-task models,  both the multi-task setting and SMILES embeddings contribute to the marked improvement in AUC-ROC  on the clinical platform while providing comparable performance on the \textit{in vitro} platform and no improvement on the \textit{in vivo} platform (Fig~\ref{fig:auc-roc-moleculenet-embed}, Section~\ref{auc}). Regarding the balanced accuracy, the multi-task model improved on the clinical platform and showed no improvement on the \textit{in vitro} or \textit{in vivo} platforms (Fig~\ref{fig:test-ba}, Section~\ref{ba}). 

The improvement in AUC-ROC correlates with the size of the datasets. ClinTox is the smallest dataset ($\sim$ 1,000 compounds), followed by Tox21 ($\sim$ 8,000 compounds) and RTECS ($\sim$ 40,000 compounds). It is possible that leveraging  the relationships among chemicals from a large corpus and among \textit{in vitro}, \textit{in vivo} and clinical tasks improves learning on smaller datasets but not on larger datasets. Additionally, the poor performance in RTECS could result from the distinct nature  of chemicals within RTECS, as displayed by the lack of overlap of RTECS chemicals with  ClinTox or Tox21 chemicals in the \textit{t}-SNE mapping (Section~\ref{invivo}, Fig~\ref{fig:tsne}). A multi-task setting does not appear to help  with a larger dataset of dissimilar chemicals but does help for the smaller clinical toxicity dataset. 

The top-performing baseline model from MoleculeNet benchmark for each platform (Weave on ClinTox, GC on Tox21, IRV on RTECS) either considered graph neural nets or an ensemble of models. These MoleculeNet models were optimized via additional hyperparameter tuning. In contrast, without any additional hyperparameter tuning, leveraging bond connectivity information, or ensemble modeling,  our multi-task deep predictive model provided  reasonable accuracy across all platforms, even on RTECS (AUC-ROC of 0.78 $\pm$ 0.02 with SMILES embeddings). 

The ability to predict accurately and consistently across the \textit{in vitro}, \textit{in vivo}, and clinical platforms is important. Currently, predictive toxicology models primarily focus on predicting within one platform (Fig S1 in Supplementary). Even though these toxicities are measured at differing granularities, the relationships among these different platforms might not be apparent if modeled separately, e.g., performance on the ClinTox task improved using the multi-task model (MTDNN-SE) as compared to the single-task models (STDNN-FP and STDNN-FE) (Fig~\ref{fig:auc-roc-moleculenet-embed}, Section~\ref{auc}). A multi-task model can overcome the small overlap in common chemicals across these platforms by sharing weights while training. For instance, even with the small overlap of the RTECS dataset with ClinTox or Tox21, the MTDNN-SE could still reasonably predict the RTECS endpoints. Predicting across these platforms together also provides a methodology to test the ability of a particular type of platform to predict clinical toxicity. 

Using our multi-task model and its transfer learning counterpart, we demonstrated the minimal relative importance of \textit{in vivo} data to make accurate predictions of clinical toxicity. The addition of \textit{in vivo} data in the MTDNN or its transfer learning counterpart did not improve clinical toxicity (Section~\ref{invivo}). Instead, the addition of \textit{in vitro} data to clinical data was sufficient in improving the predictions of clinical toxicity by AUC-ROC (Fig~\ref{fig:auc-roc-moleculenet-embed}, Section~\ref{auc}). In vivo data only helped increase balanced accuracy on the clinical task when the MTDNN was trained on Morgan fingerprints as input (light red versus light purple or light orange in Fig 3), but not when trained on SMILES embeddings (dark purple versus dark red or dark orange in Fig 3). Thus with the use of SMILES embeddings as input, in vivo data is not needed for high AUC-ROC and balanced accuracy performance on the clinical task. Moreover, the chemicals in the clinical dataset (ClinTox) were more similar to chemicals in the \textit{in vitro} dataset (Tox21) than in the \textit{in vivo} dataset (RTECS), as examined through the \textit{t}-sne mappings (Fig~\ref{fig:tsne}). The present study examines the ability of acute oral toxicity data in mice as \textit{in vivo} data, and nuclear receptor and stress response assays as \textit{in vitro} data, to predict the failure of drugs in clinical phase trials in humans. Broader \textit{in vivo} and \textit{in vitro} datasets could augment our conclusions and will be investigated in the future. 

We have improved accuracy in predicting clinical toxicity using DNN models, but with the caveat of reduced explainability in the models which we tackled in this study \cite{Jimenez-Luna2020}. DNNs are known for being "black-box" models due to their inability to describe why a prediction was made \cite{VonEschenbach2021}. Explainability of models can improve trust and adoption of models into the healthcare industry \cite{VonEschenbach2021}. Traditionally, explanations of toxicity predictions have been limited to pinpointing the \textit{presence} of substructures (Section 1.1. of Supplementary). Recent work provides molecular counterfactual explanations on toxicity\cite{numeroso2021meg} and other molecular property predictions \cite{Wellawatte2021}, using model agnostic \cite{Wellawatte2021} or reinforcement learning-based deep graph explainers \cite{numeroso2021meg}. Contrastive Explanations, provide both \textit{present} and \textit{absent} substructures correlating to a prediction, and to our knowledge has only been applied by our previous work on one specific \textit{in vitro} endpoint\cite{Lim2020}. This current work expands on current molecular explanations by providing more complete, human-understandable, contrastive explanations on toxicity predictions across \textit{in vitro}, \textit{in vivo}, and clinical platforms, with both substructures that are \textit{absent} and \textit{present} in the chemicals. For this purpose, we explain predictions made by the STDNN-FP model. We chose this model because of its consistent performance  across all tasks,  agreeing with an earlier observation that simple descriptor-based (e.g., FP) models can provide better performance and higher computational efficiency than more complex models such as the graph-based ones \cite{Jiang2021}. We explained toxicity predictions across \textit{in vitro} (Tox21), \textit{in vivo} (RTECS), and clinical (ClinTox) tasks. The PP and PN substructures obtained by the CEM not only provided explanations on toxicity predictions, but also suggested possible toxicophores and nontoxic substructures. The uniqueness of this approach is the computational identification of toxicophores from substructures absent from a molecule that would flip the prediction from nontoxic to toxic. We have thus, in the process of explaining DNN toxicity predictions, created a new and more comprehensive approach of obtaining computational toxicophores and nontoxic substructures. 

Common toxicophores were obtained across all platforms, both by PPs of "toxic" predicted molecules  and PNs of "nontoxic" predicted molecules, containing O and N groups, P, S, I and F, or aryl substructures of varying complexity with the most complex substructures present in ClinTox (Section \ref{cem}, Fig~\ref{fig:toxic}). Similar nontoxic substructures were also identified across all platforms by PP substructures of nontoxic molecules and PNs of toxic molecules, containing aryl groups, and smaller O- and N-containing substructures. The mismatch was in PNs of toxic molecules, with RTECS containing S substructures (i.e., sulfonic acid) and Tox21 containing aryl chlorides (Fig~\ref{fig:toxic}). These toxicophores identified from the CEM were verified by matching to known toxicophores, both by PP substructures \textit{present} in toxic molecules and PNs substructures \textit{absent} in nontoxic molecules (Section \ref{verification}, Fig~\ref{fig:matched}).  

The CEM-derived and verified toxicophores for the clinical task were found for both the \textit{in vitro} and \textit{in vivo} tasks, supporting the validity of initial virtual screening for known \textit{in vitro} and \textit{in vivo} toxicophores. A larger number of verified toxicophores was found for the \textit{in vitro} (Tox21) task ($\sim$ 1,000s), followed by the \textit{in vivo} (RTECS) task ($\sim$ 100s), and the last by the clinical task ($\sim$ 10s), perhaps due to the known toxicophores primarily being collected from other \textit{in vitro} experiments, as well as due to difference in dataset size. The more extensive toxicophore recovery from \textit{in vitro} and \textit{in vivo} endpoints potentially uncovers a preference in toxicophore data towards \textit{in vitro} and \textit{in vivo} experimental data. 

Computationally obtained toxicophores and nontoxic substructures from CEM explanations are bound by the three datasets used in this work. These toxicophores are most relevant to nuclear receptor and stress response assays, acute oral toxicity in mice, and toxicity in clinical phase trials. Toxicophores are used across different endpoints but mainly as an initial screening technique for potentially toxic molecules \cite{Hevener2018}. It is important to note that our toxicity explanations are an initial approach to provide more confidence in toxicity predictions made from ML/DL models and may provide an initial screening method for identifying toxic molecules, including drug candidates. The presence or absence of a toxicophore, physiologically, does not necessarily guarantee a molecule will be toxic; as potential biological targets, pathways and interactions are not taken into account in this analysis. Another limitation of our approach is only the identification of missing substructures (PNs) and not their relative location on the compound. Often carbon fragments were identified as both PP and PN substructures, due to their prevalence in Morgan Fingerprints, however more meaningful substructures were also pinpointed. The CEM is focused on explaining neural network decisions, but in future we will address comparison of explanations resulting from different predictive models and post-hoc interpretability methods. Despite these limitations, comprehensive contrastive explanations are a good starting point to illuminate "black-box" DNN-based models that have increasingly been applied to the chemical and drug toxicity predicitions while also providing a list of potential toxicophores and non-toxicophores. 

\section{Conclusion}

We have demonstrated the advantage of employing a deep neural net to predict toxicity across \textit{in vitro}, \textit{in vivo}, and clinical platforms. With pre-trained molecular SMILES embeddings as input, the multi-task model yielded improved or on par clinical toxicity predictions to current baseline and state-of-art molecular graph-based models. Unlike graph neural nets, our framework takes advantage of reduced inference costs from language models when using pre-trained SMILES embeddings. The results presented here strongly suggest that there is a minimal relative importance of \textit{in vivo} data for predicting clinical toxicity in particular when unsupervised pre-trained SMILES embeddings were used as an input to multi-task models; thus, providing possible guidance on what aspects of animal data need not be considered in predicting clinical toxicity.  We further provided a more complete and consistent molecular explanation of the predicted toxicities of a performant deep neural net across different platforms  by analyzing the contrastive substructures present within a molecule. To our knowledge, this is the first work to explain with both present \textit{and} absent substructures predictions of clinical and \textit{in vivo} toxicity. Thus we have created a framework to provide improved and explainable clinical toxicity predictions, while limiting the amount of animal data used.

\section{Methods}
\subsection{Deep Single-Task and Multi-Task Predictive Models}
\subsubsection{Input Molecular Representations} \label{input} 
Two types of computable molecular representations of chemical structures were used as input to the multi-task and single-task deep predictive models: Morgan fingerprints (FP), and pre-trained SMILES embeddings (SE). FPs, simpler and more widely used \cite{OBoyle2016}, represent molecules as a vector indicating presence of a circular substructure within varying radii around an atom. FPs were calculated by RDkit\cite{rdkit}, with a Morgan radius of 2 and bit size of 4096. SE were created in-house using a neural network-based translation model that translates from non-canonical SMILES to canonical SMILES, thereby encoding for the relationships between the chemicals. To obtain a base model for the relationships among chemicals, the translation model was trained on 103 million chemicals in Pubchem \cite{Sunghwan2020} and 35 million chemicals in ZINC-12 \cite{Irwin2012}. The trained translation model was then applied to the chemicals present in the \textit{in vitro}, \textit{in vivo}, and clinical datasets to obtain their SE. 

\subsubsection{Datasets - \textit{In vitro}, \textit{In vivo}, Clinical} \label{datasets}

The single-task and multi-task models predict binary classes, whether a chemical is toxic or not, for each of the \textit{in vitro}, \textit{in vivo}, and clinical platforms. Twelve binary classes were defined for the \textit{in vitro} platform, collected from the Tox21 challenge, a subset of the broader “Toxicology in the 21st Century” initiative that experimentally tests \textit{in vitro} the ability of a large number of chemicals to disrupt biological pathways through high-throughput screening (HTS) techniques \cite{tox21,Tice2013,Kavlock2009}. In particular, twelve \textit{in vitro} assay results were provided in Tox21, testing seven different nuclear receptor signaling effects, and five stress response effects \cite{Huang2016} of 8,014 molecules in cells \cite{Wu2017a}.

One binary class was defined for the clinical platform from the ClinTox dataset \cite{Wu2017a}, as whether a chemical was approved or failed due to toxicity in clinical phase trials. ClinTox is a curated dataset by MoleculeNet \cite{Wu2017a}, a benchmark for molecular machine learning models specifying datasets, models and evaluation criteria. ClinTox contains  1491 drugs that have either been approved by the  FDA (collected from the SWEETLEAD database \cite{Novick2013}) or failed clinical trials as reported by the Aggregate Analysis of ClinicalTrials.gov (AACT) dataset \cite{AACT}. 

One \textit{in vivo} class was parsed from the commercially available RTECS (Registry of Toxic Effects of Chemical Substances) dataset. This dataset contains \textit{in vivo} toxicity data curated from literature across various endpoints (acute, mutation, reproductive, irritation, tumorigenic, multiple-dose toxicities) in the form of different toxic measurements.  We focused on acute oral toxicity in mice due to the largest number of examples. For 42,639 chemicals, binary class was defined by LD\textsubscript{50} (lethal dose for 50 percent of the population) cutoff of 5,000 mg/kg ($\leq$ as toxic,  $>$ as nontoxic) as specified by EPA (Environmental Protection Agency) and GHS (The Globally Harmonized System of Classification and Labeling of Chemicals). 

\subsubsection{Multi-task and Single-task DNN Architecture} 

The Multi-task Deep Neural Network (MTDNN) consists of an input layer of either Morgan fingerprints (radius=2) or SMILES embeddings, passed to two layers (2048, 1024 nodes) shared by all tasks, and further two layers (512, 256 nodes) for each separate task. The output layer (one node) corresponds to toxic/nontoxic labels for each endpoint and is activated by a sigmoid (Fig S3 in Supplementary). The model is trained on random batches of 512 per training step. The number of epochs is set at the lowest validation loss (one to four epochs depending on the seed). Binary cross entropy was chosen as the loss function, and it was optimized using the Adam optimizer with a learning rate of 0.001. 

The Single-task Deep Neural Network (STDNN) consists of two layers (512, 256 nodes) and one output layer (one node) activated by a sigmoid. For tasks with multiple classes, the average of the area under the receiver operating characteristic curve (AUC-ROC) and balanced accuracy was taken.  The same data splits, and training hyperparameters are used as the MTDNN. The STDNN and MTDNN models were run with NVIDIA k80 GPUs on the Cognitive Computing Clusters at IBM.

To obtain baseline results from MoleculeNet on RTECS, the provided benchmark methods in the DeepChem \cite{Ramsundar-et-al-2019} MoleculeNet Github (https://github.com/deepchem/deepchem/tree/master/deepchem/molnet) was adapted to train and test the same models run on ClinTox and Tox21\cite{Wu2018}, on RTECS. The MoleculeNet models tested were: Weave, GC, Bypass, Multitask, IRV, RF, XGBoost, KernelSVM, and Logreg. Wu et al.\cite{Wu2018} provide details on these models.

AUC-ROC and balanced accuracy were used as metrics of performance for the models. Balanced accuracy is the average of sensitivity and specificity. The former is the fraction of
correctly classified positive classes out of all possible positives in the dataset, i.e., fraction of true positives out of correctly classified positives and falsely classified negatives, such that, 
  
\begin{equation} 
\begin{aligned}
    Sensitivity = \frac{TP}{TP + FN}.
 \end{aligned}
\end{equation}

Conversely, the specificity is this measure for the true negatives of the model, i.e. the fraction of true negatives correctly classified out of the total number of negatives in the dataset (both the correctly classified negatives and the falsely classified positives), such that, 

\begin{equation} 
\begin{aligned}
    Specificity = \frac{TN}{TN + FP}.
\end{aligned}
\end{equation}

Thus, balanced accuracy takes into account both the true negative and true positive distribution in the overall performance of model, such that, 

\begin{equation} 
\begin{aligned}
    Balanced\ Accuracy = \frac{Sensitivity + Specificity}{2}.
\end{aligned}
\end{equation}

\subsection{Transfer Learning} \label{transfer}

The base model was trained with the multi-task DNN on different combinations of \textit{in vitro} (Tox21) and \textit{in vivo} (RTECS) tasks for 2-8 epochs depending on the lowest validation loss. From the base multi-task DNN model, the two layers shared among the tasks were extracted, and their weights frozen To this, two additional layers were added for the ClinTox task on which transfer learning was needed. The number of epochs on this additional training with the transfer learning task was varied; one epoch was chosen to limit the training on the transferred task.. The same ClinTox, Tox21, and RTECS testing data as the MTDNN were used to evaluate the models.

\subsection{\textit{t}-distributed stochastic neighbor embedding} \label{tsne}

\textit{t}-Distributed stochastic neighbor embedding or \textit{t}-SNE \cite{tsne} visualizes high-dimensional data by mapping them to a low-dimensional space indicating the similarities between different points. We visualized SMILES embeddings of the chemicals present in the Tox21, RTECS, and ClinTox datasets through a \textit{t}-SNE set at 40 perplexity, 5000 iterations, and 5000 learning rates. \textit{t}-sne method in the sckit-learn package was used \cite{scikit-learn}.

\subsection{Contrastive Explanations Method} \label{CEM}
The Contrastive Explanations Method~\cite{Dhurandhar2018} provides  explanations to rationalize the classification of an input by identifying the \textit{minimal} and \textit{necessary} features which are both \textit{present} and \textit{absent} for a particular classification. The CEM thus introduces the notion of Pertinent Negative (PN) and Pertinent Positive (PP). A PN is a subset of the feature set necessary for a classifier to predict a given class, while a PP is the minimal subset of features whose presence gives rise to its prediction.

The CEM obtains the PN and PP via an optimization problem to look for a required minimum perturbation to the model, using a projected fast iterative shrinkage-thresholding algorithm (FISTA) \cite{Dhurandhar2018}. A given input example $(x_0, t_0)$ is perturbed, with $x_0$ belonging to data space $\chi$ $(x_0 \in \chi)$ and its class label $t_0$ predicted from a neural network model. The perturbed example $(x \in \chi)$ is given by $x = x_o + \delta$, with $\delta$ defining the perturbation. The PP and PNs are obtained by optimizing on this $\delta$ perturbation. Further, an autoencoder $AE(\cdot)$ is used to assure closeness of the perturbed example $x$ to the original example $x_0$, with $AE(x)$ defined at the autoencoder reconstructed example $x$. The optimization problem for obtaining the PN of $(x_0, t_0)$ is given by, 

\begin{equation}
\begin{aligned}
& \underset{\delta \in \chi \mathbin{/} x_0 }{\text{min}}
& & c \cdot f_\kappa^{neg}(x_o, \delta) + \beta \norm{\delta}_1 + \norm{\delta}_2^2 +  \gamma \norm{x_0 + \delta - AE(x_0 + \delta)}_2^2. 
\end{aligned}
\end{equation}

While the optimization problem for obtaining PP for the given example is,
\begin{equation}
\begin{aligned}
& \underset{\delta \in \chi \cap x_0 }{\text{min}}
& & c \cdot f_\kappa^{pos}(x_o, \delta) + \beta \norm{\delta}_1 + \norm{\delta}_2^2 +  \gamma \norm{\delta - AE(\delta)}_2^2. 
\end{aligned}
\end{equation}

Here, $\kappa$ defines the minimum confidence gap between the changed class probability and the original class probability, $\beta$ controls the sparsity of the solution, and $\gamma$ controls the degree of adherence to an additional autoencoder. Refer to Dhurandhar et al.\cite{Dhurandhar2018} for more details. Code is provided at https://github.com/IBM/Contrastive-Explanation-Method. 

For Tox21, ClinTox and RTECS, $\kappa$ of 0.01 and $\beta$ of 0.99 was used for PNs, while $\kappa$ was set at 0.01 for PPs. $\beta$ for PPs was set at the minimum possible to obtain a substructure: 0.1 for ClinTox and RTECS, and 0.31 for Tox21. $\gamma$ was 0 for both PPs and PNs as an additional adherence to an autoencoder was not used. Maximum of 1000 iterations were allowed, with an initial coefficient of 10 used for the main loss term and permitting nine updates to this coefficient. The seed was set at 122.

As verification, toxicophores obtained by the CEM were matched to known toxicophores. Known toxicophores were curated as known mutagenic toxicophores from \textit{in vitro} data \cite{Kazius2005} and computational models \cite{Yang2017}, or as commonly filtered reactive substructures \cite{Hevener2018}.

\section{Acknowledgements}

The authors acknowledge the support received from People for Animals Chennai Charitable Trust- an independent animal charity -for making available the RTECS database for the study. The authors also thank IBM Science for Social Good team for support. The project was conceived from conversations between S.P., the IBM Science from Social Good team, and authors at RPI who envisioned the use of Machine learning in a project called D.O.N.T (Dogs are not for Testing) to replace dogs in toxicity testing.

\section{Data availability statement}

The datasets, Tox21 and ClinTox are publicly available (https://github.com/deepchem/deepchem/tree/master/deepchem/molnet). RTECS® is commercially available from Biovia.

\bibliography{biblio}

\begin{thebibliography}{10}
\urlstyle{rm}
\expandafter\ifx\csname url\endcsname\relax
  \def\url#1{\texttt{#1}}\fi
\expandafter\ifx\csname urlprefix\endcsname\relax\def\urlprefix{URL }\fi
\expandafter\ifx\csname doiprefix\endcsname\relax\def\doiprefix{DOI: }\fi
\providecommand{\bibinfo}[2]{#2}
\providecommand{\eprint}[2][]{\url{#2}}

\bibitem{Hwang2016}
\bibinfo{author}{Hwang, T.~J.} \emph{et~al.}
\newblock \bibinfo{journal}{\bibinfo{title}{Failure of investigational drugs in
  late-stage clinical development and publication of trial results}}.
\newblock {\emph{\JournalTitle{JAMA Internal Medicine}}}
  \textbf{\bibinfo{volume}{176}}, \bibinfo{pages}{1826--1833}
  (\bibinfo{year}{2016}).

\bibitem{Hay2014}
\bibinfo{author}{Hay, M.}, \bibinfo{author}{Thomas, D.~W.},
  \bibinfo{author}{Craighead, J.~L.}, \bibinfo{author}{Economides, C.} \&
  \bibinfo{author}{Rosenthal, J.}
\newblock \bibinfo{journal}{\bibinfo{title}{Clinical development success rates
  for investigational drugs}}.
\newblock {\emph{\JournalTitle{Nature Biotechnology}}}
  \textbf{\bibinfo{volume}{32}}, \bibinfo{pages}{40--51}
  (\bibinfo{year}{2014}).

\bibitem{chenthamarakshan2020cogmol}
\bibinfo{author}{Chenthamarakshan, V.} \emph{et~al.}
\newblock \bibinfo{title}{{CogMol: Target-Specific and Selective Drug Design
  for COVID-19 Using Deep Generative Models}}.
\newblock In \bibinfo{editor}{Larochelle, H.}, \bibinfo{editor}{Ranzato, M.},
  \bibinfo{editor}{Hadsell, R.}, \bibinfo{editor}{Balcan, M.~F.} \&
  \bibinfo{editor}{Lin, H.} (eds.) \emph{\bibinfo{booktitle}{Advances in Neural
  Information Processing Systems}}, vol.~\bibinfo{volume}{33},
  \bibinfo{pages}{4320--4332} (\bibinfo{publisher}{Curran Associates, Inc.},
  \bibinfo{year}{2020}).

\bibitem{Zhavoronkov2019}
\bibinfo{author}{Zhavoronkov, A.} \emph{et~al.}
\newblock \bibinfo{journal}{\bibinfo{title}{{Deep learning enables rapid
  identification of potent DDR1 kinase inhibitors}}}.
\newblock {\emph{\JournalTitle{Nature Biotechnology}}}
  \textbf{\bibinfo{volume}{37}}, \bibinfo{pages}{1038--1040},
  \doiprefix\url{10.1038/s41587-019-0224-x} (\bibinfo{year}{2019}).

\bibitem{Li2018}
\bibinfo{author}{Li, Y.}, \bibinfo{author}{Zhang, L.} \& \bibinfo{author}{Liu,
  Z.}
\newblock \bibinfo{journal}{\bibinfo{title}{{Multi-objective de novo drug
  design with conditional graph generative model}}}.
\newblock {\emph{\JournalTitle{Journal of Cheminformatics}}}
  \textbf{\bibinfo{volume}{10}}, \bibinfo{pages}{33},
  \doiprefix\url{10.1186/s13321-018-0287-6} (\bibinfo{year}{2018}).
\newblock \eprint{1801.07299}.

\bibitem{Luco1997}
\bibinfo{author}{Luco, J.~M.} \& \bibinfo{author}{Ferretti, F.~H.}
\newblock \bibinfo{journal}{\bibinfo{title}{{QSAR} based on multiple linear
  regression and {PLS} methods for the {anti-HIV} activity of a large group of
  {HEPT} derivatives}}.
\newblock {\emph{\JournalTitle{Journal of Chemical Information and Computer
  Sciences}}} \textbf{\bibinfo{volume}{37}}, \bibinfo{pages}{392--401}
  (\bibinfo{year}{1997}).

\bibitem{Abdelaziz2016}
\bibinfo{author}{Abdelaziz, A.}, \bibinfo{author}{Spahn-Langguth, H.},
  \bibinfo{author}{Schramm, K.-W.} \& \bibinfo{author}{Tetko, I.~V.}
\newblock \bibinfo{journal}{\bibinfo{title}{{Consensus Modeling for HTS Assays
  Using In silico Descriptors Calculates the Best Balanced Accuracy in Tox21
  Challenge}}}.
\newblock {\emph{\JournalTitle{Frontiers in Environmental Science}}}
  \textbf{\bibinfo{volume}{4}}, \bibinfo{pages}{2},
  \doiprefix\url{10.3389/fenvs.2016.00002} (\bibinfo{year}{2016}).

\bibitem{Mayr2016}
\bibinfo{author}{Mayr, A.}, \bibinfo{author}{Klambauer, G.},
  \bibinfo{author}{Unterthiner, T.} \& \bibinfo{author}{Hochreiter, S.}
\newblock \bibinfo{journal}{\bibinfo{title}{{DeepTox: Toxicity Prediction using
  Deep Learning}}}.
\newblock {\emph{\JournalTitle{Frontiers in Environmental Science}}}
  \textbf{\bibinfo{volume}{3}}, \bibinfo{pages}{80},
  \doiprefix\url{10.3389/fenvs.2015.00080} (\bibinfo{year}{2016}).

\bibitem{Matsuzaka2019}
\bibinfo{author}{Matsuzaka, Y.} \& \bibinfo{author}{Uesawa, Y.}
\newblock \bibinfo{journal}{\bibinfo{title}{Prediction model with
  high-performance constitutive androstane receptor {(CAR)} using
  {DeepSnap-Deep Learning} approach from the {Tox21} {10K} compound library}}.
\newblock {\emph{\JournalTitle{International Journal of Molecular Sciences}}}
  \textbf{\bibinfo{volume}{20}}, \bibinfo{pages}{4855} (\bibinfo{year}{2019}).

\bibitem{Fernandez2018}
\bibinfo{author}{Fernandez, M.} \emph{et~al.}
\newblock \bibinfo{journal}{\bibinfo{title}{Toxic colors: the use of deep
  learning for predicting toxicity of compounds merely from their graphic
  images}}.
\newblock {\emph{\JournalTitle{Journal of Chemical Information and Modeling}}}
  \textbf{\bibinfo{volume}{58}}, \bibinfo{pages}{1533--1543}
  (\bibinfo{year}{2018}).

\bibitem{Ajmani2006}
\bibinfo{author}{Ajmani, S.}, \bibinfo{author}{Jadhav, K.} \&
  \bibinfo{author}{Kulkarni, S.~A.}
\newblock \bibinfo{journal}{\bibinfo{title}{Three-dimensional {QSAR} using the
  k-nearest neighbor method and its interpretation}}.
\newblock {\emph{\JournalTitle{Journal of Chemical Information and Modeling}}}
  \textbf{\bibinfo{volume}{46}}, \bibinfo{pages}{24--31}
  (\bibinfo{year}{2006}).

\bibitem{Chavan2015}
\bibinfo{author}{Chavan, S.}, \bibinfo{author}{Friedman, R.} \&
  \bibinfo{author}{Nicholls, I.~A.}
\newblock \bibinfo{journal}{\bibinfo{title}{Acute toxicity-supported chronic
  toxicity prediction: a k-nearest neighbor coupled read-across strategy}}.
\newblock {\emph{\JournalTitle{International Journal of Molecular Sciences}}}
  \textbf{\bibinfo{volume}{16}}, \bibinfo{pages}{11659--11677}
  (\bibinfo{year}{2015}).

\bibitem{Cao2012}
\bibinfo{author}{Cao, D.-S.} \emph{et~al.}
\newblock \bibinfo{journal}{\bibinfo{title}{In silico toxicity prediction by
  support vector machine and {SMILES} representation-based string kernel}}.
\newblock {\emph{\JournalTitle{SAR and QSAR in Environmental Research}}}
  \textbf{\bibinfo{volume}{23}}, \bibinfo{pages}{141--153}
  (\bibinfo{year}{2012}).

\bibitem{Polishchuk2009}
\bibinfo{author}{Polishchuk, P.~G.} \emph{et~al.}
\newblock \bibinfo{journal}{\bibinfo{title}{Application of random forest
  approach to {QSAR} prediction of aquatic toxicity}}.
\newblock {\emph{\JournalTitle{Journal of Chemical Information and Modeling}}}
  \textbf{\bibinfo{volume}{49}}, \bibinfo{pages}{2481--2488}
  (\bibinfo{year}{2009}).

\bibitem{Jimenez-Carretero2018}
\bibinfo{author}{Jimenez-Carretero, D.} \emph{et~al.}
\newblock \bibinfo{journal}{\bibinfo{title}{{Tox\_(R)CNN}: Deep learning-based
  nuclei profiling tool for drug toxicity screening}}.
\newblock {\emph{\JournalTitle{PLoS Computational Biology}}}
  \textbf{\bibinfo{volume}{14}} (\bibinfo{year}{2018}).

\bibitem{Huang2016}
\bibinfo{author}{Huang, R.} \emph{et~al.}
\newblock \bibinfo{journal}{\bibinfo{title}{{Tox21} challenge to build
  predictive models of nuclear receptor and stress response pathways as
  mediated by exposure to environmental chemicals and drugs}}.
\newblock {\emph{\JournalTitle{Frontiers in Environmental Science}}}
  \textbf{\bibinfo{volume}{3}}, \bibinfo{pages}{85} (\bibinfo{year}{2016}).

\bibitem{tox21}
\bibinfo{author}{Krewski, D.} \emph{et~al.}
\newblock \bibinfo{journal}{\bibinfo{title}{Toxicity testing in the 21st
  century: A vision and a strategy}}.
\newblock {\emph{\JournalTitle{Journal of Toxicology and Environmental Health,
  Part B}}} \textbf{\bibinfo{volume}{13}}, \bibinfo{pages}{51--138}
  (\bibinfo{year}{2010}).

\bibitem{Tice2013}
\bibinfo{author}{Tice, R.~R.}, \bibinfo{author}{Austin, C.~P.},
  \bibinfo{author}{Kavlock, R.~J.} \& \bibinfo{author}{Bucher, J.~R.}
\newblock \bibinfo{journal}{\bibinfo{title}{Improving the human hazard
  characterization of chemicals: a {Tox21} update}}.
\newblock {\emph{\JournalTitle{Environmental Health Perspectives}}}
  \textbf{\bibinfo{volume}{121}}, \bibinfo{pages}{756--765}
  (\bibinfo{year}{2013}).

\bibitem{Kavlock2009}
\bibinfo{author}{Kavlock, R.~J.}, \bibinfo{author}{Austin, C.~P.} \&
  \bibinfo{author}{Tice, R.~R.}
\newblock \bibinfo{journal}{\bibinfo{title}{Toxicity testing in the 21st
  century: Implications for human health risk assessment}}.
\newblock {\emph{\JournalTitle{Risk Analysis}}} \textbf{\bibinfo{volume}{29}},
  \bibinfo{pages}{485--487} (\bibinfo{year}{2009}).

\bibitem{Altae-Tran2017}
\bibinfo{author}{Altae-Tran, H.}, \bibinfo{author}{Ramsundar, B.},
  \bibinfo{author}{Pappu, A.~S.} \& \bibinfo{author}{Pande, V.}
\newblock \bibinfo{journal}{\bibinfo{title}{{Low Data Drug Discovery with
  One-Shot Learning}}}.
\newblock {\emph{\JournalTitle{ACS Central Science}}}
  \textbf{\bibinfo{volume}{3}}, \bibinfo{pages}{283--293},
  \doiprefix\url{10.1021/acscentsci.6b00367} (\bibinfo{year}{2017}).
\newblock \eprint{1611.03199}.

\bibitem{Cox2016}
\bibinfo{author}{Cox, L.~A.}, \bibinfo{author}{Popken, D.~A.},
  \bibinfo{author}{Kaplan, A.~M.}, \bibinfo{author}{Plunkett, L.~M.} \&
  \bibinfo{author}{Becker, R.~A.}
\newblock \bibinfo{journal}{\bibinfo{title}{{How well can in vitro data predict
  in vivo effects of chemicals? Rodent carcinogenicity as a case study}}}.
\newblock {\emph{\JournalTitle{Regulatory Toxicology and Pharmacology}}}
  \textbf{\bibinfo{volume}{77}}, \bibinfo{pages}{54--64},
  \doiprefix\url{10.1016/j.yrtph.2016.02.005} (\bibinfo{year}{2016}).

\bibitem{Otava2015}
\bibinfo{author}{Otava, M.}, \bibinfo{author}{Shkedy, Z.},
  \bibinfo{author}{Talloen, W.}, \bibinfo{author}{Verheyen, G.~R.} \&
  \bibinfo{author}{Kasim, A.}
\newblock \bibinfo{journal}{\bibinfo{title}{{Identification of in vitro and in
  vivo disconnects using transcriptomic data}}}.
\newblock {\emph{\JournalTitle{BMC Genomics}}} \textbf{\bibinfo{volume}{16}},
  \bibinfo{pages}{1--10}, \doiprefix\url{10.1186/s12864-015-1726-7}
  (\bibinfo{year}{2015}).

\bibitem{Olson2000}
\bibinfo{author}{Olson, H.} \emph{et~al.}
\newblock \bibinfo{journal}{\bibinfo{title}{Concordance of the toxicity of
  pharmaceuticals in humans and in animals.}}
\newblock {\emph{\JournalTitle{Regul Toxicol Pharmacol}}}
  \textbf{\bibinfo{volume}{32}}, \bibinfo{pages}{56--67},
  \doiprefix\url{10.1006/rtph.2000.1399} (\bibinfo{year}{2000}).

\bibitem{Martin2012}
\bibinfo{author}{Martin, P.~L.} \& \bibinfo{author}{Bugelski, P.~J.}
\newblock \bibinfo{title}{{Concordance of preclinical and clinical pharmacology
  and toxicology of monoclonal antibodies and fusion proteins: Soluble
  targets}}, \doiprefix\url{10.1111/j.1476-5381.2011.01812.x}
  (\bibinfo{year}{2012}).

\bibitem{Tamaki2013}
\bibinfo{author}{Tamaki, C.} \emph{et~al.}
\newblock \bibinfo{journal}{\bibinfo{title}{{Potentials and limitations of
  nonclinical safety assessment for predicting clinical adverse drug reactions:
  correlation analysis of 142 approved drugs in Japan}}}.
\newblock {\emph{\JournalTitle{The Journal of Toxicological Sciences}}}
  \textbf{\bibinfo{volume}{38}}, \bibinfo{pages}{581--598},
  \doiprefix\url{10.2131/jts.38.581} (\bibinfo{year}{2013}).

\bibitem{Becker2017}
\bibinfo{author}{Becker, R.~A.} \emph{et~al.}
\newblock \bibinfo{journal}{\bibinfo{title}{{How well can carcinogenicity be
  predicted by high throughput “characteristics of carcinogens” mechanistic
  data?}}}
\newblock {\emph{\JournalTitle{Regulatory Toxicology and Pharmacology}}}
  \textbf{\bibinfo{volume}{90}}, \bibinfo{pages}{185--196},
  \doiprefix\url{10.1016/j.yrtph.2017.08.021} (\bibinfo{year}{2017}).

\bibitem{Liua}
\bibinfo{author}{Liu, J.}, \bibinfo{author}{Patlewicz, G.},
  \bibinfo{author}{Williams, A.~J.}, \bibinfo{author}{Thomas, R.~S.} \&
  \bibinfo{author}{Shah, I.}
\newblock \bibinfo{journal}{\bibinfo{title}{{Predicting Organ Toxicity Using in
  Vitro Bioactivity Data and Chemical Structure}}}.
\newblock {\emph{\JournalTitle{Chemical Research in Toxicology}}}
  \textbf{\bibinfo{volume}{30}}, \bibinfo{pages}{2046--2059},
  \doiprefix\url{10.1021/acs.chemrestox.7b00084} (\bibinfo{year}{2017}).

\bibitem{Gadaleta2019}
\bibinfo{author}{Gadaleta, D.} \emph{et~al.}
\newblock \bibinfo{journal}{\bibinfo{title}{{SAR and QSAR modeling of a large
  collection of LD50 rat acute oral toxicity data}}}.
\newblock {\emph{\JournalTitle{Journal of Cheminformatics}}}
  \textbf{\bibinfo{volume}{11}}, \bibinfo{pages}{58},
  \doiprefix\url{10.1186/s13321-019-0383-2} (\bibinfo{year}{2019}).

\bibitem{Li2014}
\bibinfo{author}{Li, X.} \emph{et~al.}
\newblock \bibinfo{journal}{\bibinfo{title}{{In Silico Prediction of Chemical
  Acute Oral Toxicity Using Multi-Classification Methods}}}.
\newblock {\emph{\JournalTitle{Journal of Chemical Information and Modeling}}}
  \textbf{\bibinfo{volume}{54}}, \bibinfo{pages}{1061--1069},
  \doiprefix\url{10.1021/ci5000467} (\bibinfo{year}{2014}).

\bibitem{Idakwo2019}
\bibinfo{author}{Idakwo, G.} \emph{et~al.}
\newblock \bibinfo{journal}{\bibinfo{title}{{Deep Learning-Based
  Structure-Activity Relationship Modeling for Multi-Category Toxicity
  Classification: A Case Study of 10K Tox21 Chemicals With High-Throughput
  Cell-Based Androgen Receptor Bioassay Data}}}.
\newblock {\emph{\JournalTitle{Frontiers in Physiology}}}
  \textbf{\bibinfo{volume}{10}}, \doiprefix\url{10.3389/fphys.2019.01044}
  (\bibinfo{year}{2019}).

\bibitem{Chen2013}
\bibinfo{author}{Chen, L.} \emph{et~al.}
\newblock \bibinfo{journal}{\bibinfo{title}{{Predicting Chemical Toxicity
  Effects Based on Chemical-Chemical Interactions}}}.
\newblock {\emph{\JournalTitle{PLoS ONE}}} \textbf{\bibinfo{volume}{8}},
  \bibinfo{pages}{e56517}, \doiprefix\url{10.1371/journal.pone.0056517}
  (\bibinfo{year}{2013}).

\bibitem{Jiang2015}
\bibinfo{author}{Jiang, Z.}, \bibinfo{author}{Xu, R.} \& \bibinfo{author}{Dong,
  C.}
\newblock \bibinfo{journal}{\bibinfo{title}{{Identification of Chemical
  Toxicity Using Ontology Information of Chemicals}}}.
\newblock {\emph{\JournalTitle{Computational and Mathematical Methods in
  Medicine}}} \textbf{\bibinfo{volume}{2015}},
  \doiprefix\url{10.1155/2015/246374} (\bibinfo{year}{2015}).

\bibitem{Raies2018}
\bibinfo{author}{Raies, A.~B.} \& \bibinfo{author}{Bajic, V.~B.}
\newblock \bibinfo{journal}{\bibinfo{title}{{In silico toxicology:
  comprehensive benchmarking of multi-label classification methods applied to
  chemical toxicity data}}}.
\newblock {\emph{\JournalTitle{Wiley Interdisciplinary Reviews: Computational
  Molecular Science}}} \textbf{\bibinfo{volume}{8}},
  \doiprefix\url{10.1002/wcms.1352} (\bibinfo{year}{2018}).

\bibitem{Wu2018}
\bibinfo{author}{Wu, Z.} \emph{et~al.}
\newblock \bibinfo{journal}{\bibinfo{title}{Molecule{N}et: a benchmark for
  molecular machine learning}}.
\newblock {\emph{\JournalTitle{Chemical Science}}}
  \textbf{\bibinfo{volume}{9}}, \bibinfo{pages}{513--530}
  (\bibinfo{year}{2018}).

\bibitem{Sosnin2019}
\bibinfo{author}{Sosnin, S.}, \bibinfo{author}{Karlov, D.},
  \bibinfo{author}{Tetko, I.~V.} \& \bibinfo{author}{Fedorov, M.~V.}
\newblock \bibinfo{journal}{\bibinfo{title}{{Comparative Study of Multitask
  Toxicity Modeling on a Broad Chemical Space}}}.
\newblock {\emph{\JournalTitle{Journal of Chemical Information and Modeling}}}
  \textbf{\bibinfo{volume}{59}}, \bibinfo{pages}{1062--1072},
  \doiprefix\url{10.1021/acs.jcim.8b00685} (\bibinfo{year}{2019}).

\bibitem{Xu2017}
\bibinfo{author}{Xu, Y.}, \bibinfo{author}{Pei, J.} \& \bibinfo{author}{Lai,
  L.}
\newblock \bibinfo{journal}{\bibinfo{title}{{Deep Learning Based Regression and
  Multiclass Models for Acute Oral Toxicity Prediction with Automatic Chemical
  Feature Extraction}}}.
\newblock {\emph{\JournalTitle{Journal of Chemical Information and Modeling}}}
  \textbf{\bibinfo{volume}{57}}, \bibinfo{pages}{2672--2685},
  \doiprefix\url{10.1021/acs.jcim.7b00244} (\bibinfo{year}{2017}).
\newblock \eprint{1704.04718}.

\bibitem{Tang2018}
\bibinfo{author}{Tang, W.}, \bibinfo{author}{Chen, J.}, \bibinfo{author}{Wang,
  Z.}, \bibinfo{author}{Xie, H.} \& \bibinfo{author}{Hong, H.}
\newblock \bibinfo{journal}{\bibinfo{title}{Deep learning for predicting
  toxicity of chemicals: A mini review}}.
\newblock {\emph{\JournalTitle{Journal of Environmental Science and Health,
  Part C}}} \textbf{\bibinfo{volume}{36}}, \bibinfo{pages}{252--271}
  (\bibinfo{year}{2018}).

\bibitem{OECD_guidance}
\bibinfo{author}{OECD}.
\newblock \bibinfo{title}{Guidance document for the use of {AOPs} in developing
  {IATA}} (\bibinfo{year}{2016}).
\newblock \bibinfo{note}{Accessed online}.

\bibitem{Raies2016}
\bibinfo{author}{Raies, A.~B.} \& \bibinfo{author}{Bajic, V.~B.}
\newblock \bibinfo{journal}{\bibinfo{title}{In silico toxicology: computational
  methods for the prediction of chemical toxicity}}.
\newblock {\emph{\JournalTitle{Computational Molecular Science}}}
  \textbf{\bibinfo{volume}{6}}, \bibinfo{pages}{147--172}
  (\bibinfo{year}{2016}).

\bibitem{Sharma2017}
\bibinfo{author}{Sharma, A.~K.}, \bibinfo{author}{Srivastava, G.~N.},
  \bibinfo{author}{Roy, A.} \& \bibinfo{author}{Sharma, V.~K.}
\newblock \bibinfo{journal}{\bibinfo{title}{{ToxiM}: A toxicity prediction tool
  for small molecules developed using machine learning and chemoinformatics
  approaches}}.
\newblock {\emph{\JournalTitle{Frontiers in Pharmacology}}}
  \textbf{\bibinfo{volume}{8}}, \bibinfo{pages}{880} (\bibinfo{year}{2017}).

\bibitem{Russell1959}
\bibinfo{author}{Russell, W.} \& \bibinfo{author}{Burch, R.}
\newblock \emph{\bibinfo{title}{{The Principles of Humane Experimental
  Technique}}} (\bibinfo{publisher}{London: Methuen}, \bibinfo{year}{1959}),
  \bibinfo{edition}{238} edn.
\newblock \eprint{arXiv:1011.1669v3}.

\bibitem{Tornqvist2014}
\bibinfo{author}{T{\"{o}}rnqvist, E.} \emph{et~al.}
\newblock \bibinfo{journal}{\bibinfo{title}{{Strategic focus on 3R principles
  reveals major reductions in the use of animals in pharmaceutical toxicity
  testing}}}.
\newblock {\emph{\JournalTitle{PLoS ONE}}} \textbf{\bibinfo{volume}{9}},
  \doiprefix\url{10.1371/journal.pone.0101638} (\bibinfo{year}{2014}).

\bibitem{Dhurandhar2018}
\bibinfo{author}{Dhurandhar, A.} \emph{et~al.}
\newblock \bibinfo{title}{Explanations based on the missing: Towards
  contrastive explanations with pertinent negatives}.
\newblock In \emph{\bibinfo{booktitle}{Advances in Neural Information
  Processing Systems}}, \bibinfo{pages}{592--603} (\bibinfo{year}{2018}).

\bibitem{OBoyle2016}
\bibinfo{author}{O'Boyle, N.~M.} \& \bibinfo{author}{Sayle, R.~A.}
\newblock \bibinfo{journal}{\bibinfo{title}{{Comparing structural fingerprints
  using a literature-based similarity benchmark}}}.
\newblock {\emph{\JournalTitle{Journal of Cheminformatics}}}
  \textbf{\bibinfo{volume}{8}}, \bibinfo{pages}{36},
  \doiprefix\url{10.1186/s13321-016-0148-0} (\bibinfo{year}{2016}).

\bibitem{Wu2017a}
\bibinfo{author}{Wu, Z.} \emph{et~al.}
\newblock \bibinfo{journal}{\bibinfo{title}{{MoleculeNet: A Benchmark for
  Molecular Machine Learning}}}.
\newblock {\emph{\JournalTitle{Chemical Science}}}
  \textbf{\bibinfo{volume}{9}}, \bibinfo{pages}{513--530}
  (\bibinfo{year}{2017}).
\newblock \eprint{1703.00564}.

\bibitem{Lim2020}
\bibinfo{author}{Lim, K.~W.}, \bibinfo{author}{Sharma, B.},
  \bibinfo{author}{Das, P.}, \bibinfo{author}{Chenthamarakshan, V.} \&
  \bibinfo{author}{Dordick, J.~S.}
\newblock \bibinfo{title}{Explaining chemical toxicity using missing features},
  \doiprefix\url{10.48550/ARXIV.2009.12199} (\bibinfo{year}{2020}).

\bibitem{weave}
\bibinfo{author}{Kearnes, S.}, \bibinfo{author}{McCloskey, K.},
  \bibinfo{author}{Berndl, M.}, \bibinfo{author}{Pande, V.} \&
  \bibinfo{author}{Riley, P.}
\newblock \bibinfo{journal}{\bibinfo{title}{Molecular graph convolutions:
  moving beyond fingerprints}}.
\newblock {\emph{\JournalTitle{Journal of Computer-Aided Molecular Design}}}
  \bibinfo{pages}{1--14} (\bibinfo{year}{2016}).
\newblock \bibinfo{note}{Doi:10.1007/s10822-016-9938-8
  http://dx.doi.org/10.1007/s10822-016-9938-8}.

\bibitem{Sedykh2011}
\bibinfo{author}{Sedykh, A.} \emph{et~al.}
\newblock \bibinfo{journal}{\bibinfo{title}{{Use of in Vitro HTS-Derived
  Concentration-Response Data as Biological Descriptors Improves the Accuracy
  of QSAR Models of in Vivo Toxicity}}}.
\newblock {\emph{\JournalTitle{Environmental Health Perspectives Environ Health
  Perspect}}} \textbf{\bibinfo{volume}{119}}, \bibinfo{pages}{364--370},
  \doiprefix\url{10.1289/ehp.1002476} (\bibinfo{year}{2011}).

\bibitem{Thomas2012}
\bibinfo{author}{Thomas, R.~S.} \emph{et~al.}
\newblock \bibinfo{journal}{\bibinfo{title}{{A comprehensive statistical
  analysis of predicting in vivo hazard using high-throughput in vitro
  screening}}}.
\newblock {\emph{\JournalTitle{Toxicological Sciences}}}
  \textbf{\bibinfo{volume}{128}}, \bibinfo{pages}{398--417},
  \doiprefix\url{10.1093/toxsci/kfs159} (\bibinfo{year}{2012}).

\bibitem{Weiss2016}
\bibinfo{author}{Weiss, K.}, \bibinfo{author}{Khoshgoftaar, T.~M.} \&
  \bibinfo{author}{Wang, D.~D.}
\newblock \bibinfo{journal}{\bibinfo{title}{{A survey of transfer learning}}}.
\newblock {\emph{\JournalTitle{Journal of Big Data}}}
  \textbf{\bibinfo{volume}{3}}, \bibinfo{pages}{9},
  \doiprefix\url{10.1186/s40537-016-0043-6} (\bibinfo{year}{2016}).

\bibitem{Hinton}
\bibinfo{author}{Hinton, G.~E.} \& \bibinfo{author}{Roweis, S.}
\newblock \bibinfo{title}{Stochastic neighbor embedding}.
\newblock In \bibinfo{editor}{Becker, S.}, \bibinfo{editor}{Thrun, S.} \&
  \bibinfo{editor}{Obermayer, K.} (eds.) \emph{\bibinfo{booktitle}{Advances in
  Neural Information Processing Systems}}, vol.~\bibinfo{volume}{15}
  (\bibinfo{publisher}{MIT Press}, \bibinfo{year}{2002}).

\bibitem{Yang2017}
\bibinfo{author}{Yang, H.} \emph{et~al.}
\newblock \bibinfo{journal}{\bibinfo{title}{{Evaluation of Different Methods
  for Identification of Structural Alerts Using Chemical Ames Mutagenicity Data
  Set as a Benchmark}}}.
\newblock {\emph{\JournalTitle{Chemical Research in Toxicology}}}
  \textbf{\bibinfo{volume}{30}}, \bibinfo{pages}{1355--1364},
  \doiprefix\url{10.1021/acs.chemrestox.7b00083} (\bibinfo{year}{2017}).

\bibitem{Kazius2005}
\bibinfo{author}{Kazius, J.}, \bibinfo{author}{McGuire, R.} \&
  \bibinfo{author}{Bursi, R.}
\newblock \bibinfo{journal}{\bibinfo{title}{{Derivation and validation of
  toxicophores for mutagenicity prediction}}}.
\newblock {\emph{\JournalTitle{Journal of Medicinal Chemistry}}}
  \textbf{\bibinfo{volume}{48}}, \bibinfo{pages}{312--320},
  \doiprefix\url{10.1021/jm040835a} (\bibinfo{year}{2005}).

\bibitem{Hevener2018}
\bibinfo{author}{Hevener, K.~E.}
\newblock \bibinfo{title}{{Computational toxicology methods in chemical library
  design and high-throughput screening hit validation}}.
\newblock In \emph{\bibinfo{booktitle}{Methods in Molecular Biology}}, vol.
  \bibinfo{volume}{1800}, \bibinfo{pages}{275--285},
  \doiprefix\url{10.1007/978-1-4939-7899-1_13} (\bibinfo{publisher}{Humana
  Press Inc.}, \bibinfo{year}{2018}).

\bibitem{Jimenez-Luna2020}
\bibinfo{author}{Jim{\'{e}}nez-Luna, J.}, \bibinfo{author}{Grisoni, F.} \&
  \bibinfo{author}{Schneider, G.}
\newblock \bibinfo{title}{{Drug discovery with explainable artificial
  intelligence}}, \doiprefix\url{10.1038/s42256-020-00236-4}
  (\bibinfo{year}{2020}).
\newblock \eprint{2007.00523}.

\bibitem{VonEschenbach2021}
\bibinfo{author}{von Eschenbach, W.~J.}
\newblock \bibinfo{journal}{\bibinfo{title}{{Transparency and the Black Box
  Problem: Why We Do Not Trust AI}}}.
\newblock {\emph{\JournalTitle{Philosophy and Technology}}}
  \textbf{\bibinfo{volume}{34}}, \bibinfo{pages}{1607--1622},
  \doiprefix\url{10.1007/s13347-021-00477-0} (\bibinfo{year}{2021}).

\bibitem{numeroso2021meg}
\bibinfo{author}{Numeroso, D.} \& \bibinfo{author}{Bacciu, D.}
\newblock \bibinfo{title}{Meg: Generating molecular counterfactual explanations
  for deep graph networks} (\bibinfo{year}{2021}).
\newblock \eprint{2104.08060}.

\bibitem{Wellawatte2021}
\bibinfo{author}{Wellawatte, G.~P.}, \bibinfo{author}{Seshadri, A.} \&
  \bibinfo{author}{White, A.~D.}
\newblock \bibinfo{journal}{\bibinfo{title}{Model agnostic generation of
  counterfactual explanations for molecules}}.
\newblock {\emph{\JournalTitle{ChemRxiv}}}
  \doiprefix\url{10.26434/chemrxiv-2021-4qkg8} (\bibinfo{year}{2021}).

\bibitem{Jiang2021}
\bibinfo{author}{Jiang, D.} \emph{et~al.}
\newblock \bibinfo{journal}{\bibinfo{title}{{Could graph neural networks learn
  better molecular representation for drug discovery? A comparison study of
  descriptor-based and graph-based models}}}.
\newblock {\emph{\JournalTitle{Journal of Cheminformatics}}}
  \textbf{\bibinfo{volume}{13}}, \bibinfo{pages}{12},
  \doiprefix\url{10.1186/s13321-020-00479-8} (\bibinfo{year}{2021}).

\bibitem{rdkit}
\bibinfo{title}{Rdkit: Open-source cheminformatics}.
\newblock \bibinfo{note}{\url{http://www.rdkit.org}}.

\bibitem{Sunghwan2020}
\bibinfo{author}{Kim, S.} \emph{et~al.}
\newblock \bibinfo{journal}{\bibinfo{title}{{PubChem in 2021: new data content
  and improved web interfaces}}}.
\newblock {\emph{\JournalTitle{Nucleic Acids Research}}}
  \textbf{\bibinfo{volume}{49}}, \bibinfo{pages}{D1388--D1395},
  \doiprefix\url{10.1093/nar/gkaa971} (\bibinfo{year}{2020}).
\newblock
  \eprint{https://academic.oup.com/nar/article-pdf/49/D1/D1388/35363961/gkaa971.pdf}.

\bibitem{Irwin2012}
\bibinfo{author}{Irwin, J.~J.}, \bibinfo{author}{Sterling, T.},
  \bibinfo{author}{Mysinger, M.~M.}, \bibinfo{author}{Bolstad, E.~S.} \&
  \bibinfo{author}{Coleman, R.~G.}
\newblock \bibinfo{title}{{ZINC: A free tool to discover chemistry for
  biology}}, \doiprefix\url{10.1021/ci3001277} (\bibinfo{year}{2012}).

\bibitem{Novick2013}
\bibinfo{author}{Novick, P.~A.}, \bibinfo{author}{Ortiz, O.~F.},
  \bibinfo{author}{Poelman, J.}, \bibinfo{author}{Abdulhay, A.~Y.} \&
  \bibinfo{author}{Pande, V.~S.}
\newblock \bibinfo{journal}{\bibinfo{title}{{SWEETLEAD: an In Silico Database
  of Approved Drugs, Regulated Chemicals, and Herbal Isolates for
  Computer-Aided Drug Discovery}}}.
\newblock {\emph{\JournalTitle{PLoS ONE}}} \textbf{\bibinfo{volume}{8}},
  \bibinfo{pages}{e79568}, \doiprefix\url{10.1371/journal.pone.0079568}
  (\bibinfo{year}{2013}).

\bibitem{AACT}
\bibinfo{author}{ClinicalTrials.gov}.
\newblock \bibinfo{title}{Aggregate analysis of clincaltrials.gov (aact)
  database}.
\newblock
  \bibinfo{note}{\url{http://www.ctti-clinicaltrials.org/aact-database},
  accessed 2020-06-19.}

\bibitem{Ramsundar-et-al-2019}
\bibinfo{author}{Ramsundar, B.}, \bibinfo{author}{Eastman, P.},
  \bibinfo{author}{Walters, P.} \& \bibinfo{author}{Pande, V.}
\newblock \emph{\bibinfo{title}{Deep Learning for the Life Sciences: applying
  deep learning to genomics, microscopy, drug discovery, and more}}
  (\bibinfo{publisher}{O'Reilly Media}, \bibinfo{year}{2019}).

\bibitem{tsne}
\bibinfo{author}{Maaten, L. V.~D.} \& \bibinfo{author}{Hinton, G.}
\newblock \bibinfo{journal}{\bibinfo{title}{Visualizing data using t-sne}}.
\newblock {\emph{\JournalTitle{Journal of Machine Learning Research}}}
  \textbf{\bibinfo{volume}{9}}, \bibinfo{pages}{2579--2605}
  (\bibinfo{year}{2008}).

\bibitem{scikit-learn}
\bibinfo{author}{Pedregosa, F.} \emph{et~al.}
\newblock \bibinfo{journal}{\bibinfo{title}{Scikit-learn: Machine learning in
  {P}ython}}.
\newblock {\emph{\JournalTitle{Journal of Machine Learning Research}}}
  \textbf{\bibinfo{volume}{12}}, \bibinfo{pages}{2825--2830}
  (\bibinfo{year}{2011}).

\end{thebibliography}


\begin{thebibliography}{10}
\urlstyle{rm}
\expandafter\ifx\csname url\endcsname\relax
  \def\url#1{\texttt{#1}}\fi
\expandafter\ifx\csname urlprefix\endcsname\relax\def\urlprefix{URL }\fi
\expandafter\ifx\csname doiprefix\endcsname\relax\def\doiprefix{DOI: }\fi
\providecommand{\bibinfo}[2]{#2}
\providecommand{\eprint}[2][]{\url{#2}}

\bibitem{Gadaleta2019}
\bibinfo{author}{Gadaleta, D.} \emph{et~al.}
\newblock \bibinfo{journal}{\bibinfo{title}{{SAR and QSAR modeling of a large
  collection of LD50 rat acute oral toxicity data}}}.
\newblock {\emph{\JournalTitle{Journal of Cheminformatics}}}
  \textbf{\bibinfo{volume}{11}}, \bibinfo{pages}{58},
  \doiprefix\url{10.1186/s13321-019-0383-2} (\bibinfo{year}{2019}).

\bibitem{Li2014}
\bibinfo{author}{Li, X.} \emph{et~al.}
\newblock \bibinfo{journal}{\bibinfo{title}{{In Silico Prediction of Chemical
  Acute Oral Toxicity Using Multi-Classification Methods}}}.
\newblock {\emph{\JournalTitle{Journal of Chemical Information and Modeling}}}
  \textbf{\bibinfo{volume}{54}}, \bibinfo{pages}{1061--1069},
  \doiprefix\url{10.1021/ci5000467} (\bibinfo{year}{2014}).

\bibitem{Idakwo2019}
\bibinfo{author}{Idakwo, G.} \emph{et~al.}
\newblock \bibinfo{journal}{\bibinfo{title}{{Deep Learning-Based
  Structure-Activity Relationship Modeling for Multi-Category Toxicity
  Classification: A Case Study of 10K Tox21 Chemicals With High-Throughput
  Cell-Based Androgen Receptor Bioassay Data}}}.
\newblock {\emph{\JournalTitle{Frontiers in Physiology}}}
  \textbf{\bibinfo{volume}{10}}, \doiprefix\url{10.3389/fphys.2019.01044}
  (\bibinfo{year}{2019}).

\bibitem{Chen2013}
\bibinfo{author}{Chen, L.} \emph{et~al.}
\newblock \bibinfo{journal}{\bibinfo{title}{{Predicting Chemical Toxicity
  Effects Based on Chemical-Chemical Interactions}}}.
\newblock {\emph{\JournalTitle{PLoS ONE}}} \textbf{\bibinfo{volume}{8}},
  \bibinfo{pages}{e56517}, \doiprefix\url{10.1371/journal.pone.0056517}
  (\bibinfo{year}{2013}).

\bibitem{Jiang2015}
\bibinfo{author}{Jiang, Z.}, \bibinfo{author}{Xu, R.} \& \bibinfo{author}{Dong,
  C.}
\newblock \bibinfo{journal}{\bibinfo{title}{{Identification of Chemical
  Toxicity Using Ontology Information of Chemicals}}}.
\newblock {\emph{\JournalTitle{Computational and Mathematical Methods in
  Medicine}}} \textbf{\bibinfo{volume}{2015}},
  \doiprefix\url{10.1155/2015/246374} (\bibinfo{year}{2015}).

\bibitem{Raies2018}
\bibinfo{author}{Raies, A.~B.} \& \bibinfo{author}{Bajic, V.~B.}
\newblock \bibinfo{journal}{\bibinfo{title}{{In silico toxicology:
  comprehensive benchmarking of multi-label classification methods applied to
  chemical toxicity data}}}.
\newblock {\emph{\JournalTitle{Wiley Interdisciplinary Reviews: Computational
  Molecular Science}}} \textbf{\bibinfo{volume}{8}},
  \doiprefix\url{10.1002/wcms.1352} (\bibinfo{year}{2018}).

\bibitem{Wu2018}
\bibinfo{author}{Wu, Z.} \emph{et~al.}
\newblock \bibinfo{journal}{\bibinfo{title}{Molecule{N}et: a benchmark for
  molecular machine learning}}.
\newblock {\emph{\JournalTitle{Chemical Science}}}
  \textbf{\bibinfo{volume}{9}}, \bibinfo{pages}{513--530}
  (\bibinfo{year}{2018}).

\bibitem{Sosnin2019}
\bibinfo{author}{Sosnin, S.}, \bibinfo{author}{Karlov, D.},
  \bibinfo{author}{Tetko, I.~V.} \& \bibinfo{author}{Fedorov, M.~V.}
\newblock \bibinfo{journal}{\bibinfo{title}{{Comparative Study of Multitask
  Toxicity Modeling on a Broad Chemical Space}}}.
\newblock {\emph{\JournalTitle{Journal of Chemical Information and Modeling}}}
  \textbf{\bibinfo{volume}{59}}, \bibinfo{pages}{1062--1072},
  \doiprefix\url{10.1021/acs.jcim.8b00685} (\bibinfo{year}{2019}).

\bibitem{Xu2017}
\bibinfo{author}{Xu, Y.}, \bibinfo{author}{Pei, J.} \& \bibinfo{author}{Lai,
  L.}
\newblock \bibinfo{journal}{\bibinfo{title}{{Deep Learning Based Regression and
  Multiclass Models for Acute Oral Toxicity Prediction with Automatic Chemical
  Feature Extraction}}}.
\newblock {\emph{\JournalTitle{Journal of Chemical Information and Modeling}}}
  \textbf{\bibinfo{volume}{57}}, \bibinfo{pages}{2672--2685},
  \doiprefix\url{10.1021/acs.jcim.7b00244} (\bibinfo{year}{2017}).
\newblock \eprint{1704.04718}.

\bibitem{Liu2015a}
\bibinfo{author}{Liu, J.} \emph{et~al.}
\newblock \bibinfo{journal}{\bibinfo{title}{Predicting hepatotoxicity using
  toxcast in vitro bioactivity and chemical structure}}.
\newblock {\emph{\JournalTitle{Chemical Research in Toxicology}}}
  \textbf{\bibinfo{volume}{28}}, \bibinfo{pages}{738--751},
  \doiprefix\url{10.1021/tx500501h} (\bibinfo{year}{2015}).
\newblock \bibinfo{note}{PMID: 25697799},
  \eprint{https://doi.org/10.1021/tx500501h}.

\bibitem{Liua}
\bibinfo{author}{Liu, J.}, \bibinfo{author}{Patlewicz, G.},
  \bibinfo{author}{Williams, A.~J.}, \bibinfo{author}{Thomas, R.~S.} \&
  \bibinfo{author}{Shah, I.}
\newblock \bibinfo{journal}{\bibinfo{title}{{Predicting Organ Toxicity Using in
  Vitro Bioactivity Data and Chemical Structure}}}.
\newblock {\emph{\JournalTitle{Chemical Research in Toxicology}}}
  \textbf{\bibinfo{volume}{30}}, \bibinfo{pages}{2046--2059},
  \doiprefix\url{10.1021/acs.chemrestox.7b00084} (\bibinfo{year}{2017}).

\bibitem{Abdelaziz2016}
\bibinfo{author}{Abdelaziz, A.}, \bibinfo{author}{Spahn-Langguth, H.},
  \bibinfo{author}{Schramm, K.-W.} \& \bibinfo{author}{Tetko, I.~V.}
\newblock \bibinfo{journal}{\bibinfo{title}{{Consensus Modeling for HTS Assays
  Using In silico Descriptors Calculates the Best Balanced Accuracy in Tox21
  Challenge}}}.
\newblock {\emph{\JournalTitle{Frontiers in Environmental Science}}}
  \textbf{\bibinfo{volume}{4}}, \bibinfo{pages}{2},
  \doiprefix\url{10.3389/fenvs.2016.00002} (\bibinfo{year}{2016}).

\bibitem{Mayr2016}
\bibinfo{author}{Mayr, A.}, \bibinfo{author}{Klambauer, G.},
  \bibinfo{author}{Unterthiner, T.} \& \bibinfo{author}{Hochreiter, S.}
\newblock \bibinfo{journal}{\bibinfo{title}{{DeepTox: Toxicity Prediction using
  Deep Learning}}}.
\newblock {\emph{\JournalTitle{Frontiers in Environmental Science}}}
  \textbf{\bibinfo{volume}{3}}, \bibinfo{pages}{80},
  \doiprefix\url{10.3389/fenvs.2015.00080} (\bibinfo{year}{2016}).

\bibitem{Wu2017a}
\bibinfo{author}{Wu, Z.} \emph{et~al.}
\newblock \bibinfo{journal}{\bibinfo{title}{{MoleculeNet: A Benchmark for
  Molecular Machine Learning}}}.
\newblock {\emph{\JournalTitle{Chemical Science}}}
  \textbf{\bibinfo{volume}{9}}, \bibinfo{pages}{513--530}
  (\bibinfo{year}{2017}).
\newblock \eprint{1703.00564}.

\bibitem{Perez-ParrasToledano2019}
\bibinfo{author}{{P{\'{e}}rez-Parras Toledano}, J.},
  \bibinfo{author}{Garc{\'{i}}a-Pedrajas, N.} \&
  \bibinfo{author}{Cerruela-Garc{\'{i}}a, G.}
\newblock \bibinfo{journal}{\bibinfo{title}{{Multilabel and missing label
  methods for binary quantitative structure-activity relationship models: An
  application for the prediction of adverse drug reactions}}}.
\newblock {\emph{\JournalTitle{Journal of Chemical Information and Modeling}}}
  \doiprefix\url{10.1021/acs.jcim.9b00611} (\bibinfo{year}{2019}).

\bibitem{Munoz}
\bibinfo{author}{Mu{\~{n}}oz, E.}, \bibinfo{author}{Nov{\'{a}}{\v{c}}ek, V.} \&
  \bibinfo{author}{Vandenbussche, P.~Y.}
\newblock \bibinfo{journal}{\bibinfo{title}{{Facilitating prediction of adverse
  drug reactions by using knowledge graphs and multi-label learning models}}}.
\newblock {\emph{\JournalTitle{Briefings in Bioinformatics}}}
  \textbf{\bibinfo{volume}{20}}, \bibinfo{pages}{1--13},
  \doiprefix\url{10.1093/bib/bbx099} (\bibinfo{year}{2019}).

\bibitem{Wadhwa2018a}
\bibinfo{author}{Wadhwa, S.}, \bibinfo{author}{Gupta, A.},
  \bibinfo{author}{Dokania, S.}, \bibinfo{author}{Kanji, R.} \&
  \bibinfo{author}{Bagler, G.}
\newblock \bibinfo{journal}{\bibinfo{title}{{A hierarchical anatomical
  classification schema for prediction of phenotypic side effects}}}.
\newblock {\emph{\JournalTitle{PLoS ONE}}} \textbf{\bibinfo{volume}{13}},
  \doiprefix\url{10.1371/journal.pone.0193959} (\bibinfo{year}{2018}).

\bibitem{Jimenez-Luna2020}
\bibinfo{author}{Jim{\'{e}}nez-Luna, J.}, \bibinfo{author}{Grisoni, F.} \&
  \bibinfo{author}{Schneider, G.}
\newblock \bibinfo{title}{{Drug discovery with explainable artificial
  intelligence}}, \doiprefix\url{10.1038/s42256-020-00236-4}
  (\bibinfo{year}{2020}).
\newblock \eprint{2007.00523}.

\bibitem{Polishchuk2017}
\bibinfo{author}{Polishchuk, P.}
\newblock \bibinfo{title}{{Interpretation of Quantitative Structure-Activity
  Relationship Models: Past, Present, and Future}},
  \doiprefix\url{10.1021/acs.jcim.7b00274} (\bibinfo{year}{2017}).

\bibitem{Sharma2017}
\bibinfo{author}{Sharma, A.~K.}, \bibinfo{author}{Srivastava, G.~N.},
  \bibinfo{author}{Roy, A.} \& \bibinfo{author}{Sharma, V.~K.}
\newblock \bibinfo{journal}{\bibinfo{title}{{ToxiM}: A toxicity prediction tool
  for small molecules developed using machine learning and chemoinformatics
  approaches}}.
\newblock {\emph{\JournalTitle{Frontiers in Pharmacology}}}
  \textbf{\bibinfo{volume}{8}}, \bibinfo{pages}{880} (\bibinfo{year}{2017}).

\bibitem{Rasmussen2022}
\bibinfo{author}{Rasmussen, M.~H.}, \bibinfo{author}{Christensen, D.~S.} \&
  \bibinfo{author}{Jensen, J.~H.}
\newblock \bibinfo{journal}{\bibinfo{title}{Do machines dream of atoms? a
  quantitative molecular benchmark for explainable ai heatmaps}}.
\newblock {\emph{\JournalTitle{ChemRxiv}}}
  \doiprefix\url{10.26434/chemrxiv-2022-gnq3w} (\bibinfo{year}{2022}).

\bibitem{Polishchuk2013}
\bibinfo{author}{Polishchuk, P.~G.}, \bibinfo{author}{Kuz'min, V.~E.},
  \bibinfo{author}{Artemenko, A.~G.} \& \bibinfo{author}{Muratov, E.~N.}
\newblock \bibinfo{journal}{\bibinfo{title}{{Universal Approach for Structural
  Interpretation of QSAR/QSPR Models}}}.
\newblock {\emph{\JournalTitle{Molecular Informatics}}}
  \textbf{\bibinfo{volume}{32}}, \bibinfo{pages}{843--853},
  \doiprefix\url{10.1002/minf.201300029} (\bibinfo{year}{2013}).

\bibitem{Pope2019}
\bibinfo{author}{E~Pope, P.}, \bibinfo{author}{Kolouri, S.},
  \bibinfo{author}{Rostami, M.}, \bibinfo{author}{Martin, C.} \&
  \bibinfo{author}{Hoffmann, H.}
\newblock \bibinfo{title}{Explainability methods for graph convolutional neural
  networks}.
\newblock In \emph{\bibinfo{booktitle}{The IEEE Conference on Computer Vision
  and Pattern Recognition (CVPR)}}, \bibinfo{pages}{10772--10781}
  (\bibinfo{year}{2019}).

\bibitem{Akita2018}
\bibinfo{author}{Akita, H.} \emph{et~al.}
\newblock \bibinfo{title}{{BayesGrad}: Explaining predictions of graph
  convolutional networks}.
\newblock In \emph{\bibinfo{booktitle}{International Conference on Neural
  Information Processing}}, \bibinfo{pages}{81--92}
  (\bibinfo{organization}{Springer}, \bibinfo{year}{2018}).

\bibitem{Jimenez-Luna2021}
\bibinfo{author}{Jim{\'{e}}nez-Luna, J.}, \bibinfo{author}{Skalic, M.},
  \bibinfo{author}{Weskamp, N.} \& \bibinfo{author}{Schneider, G.}
\newblock \bibinfo{journal}{\bibinfo{title}{{Coloring Molecules with
  Explainable Artificial Intelligence for Preclinical Relevance Assessment}}}.
\newblock {\emph{\JournalTitle{Journal of Chemical Information and Modeling}}}
  \textbf{\bibinfo{volume}{61}}, \bibinfo{pages}{1083--1094},
  \doiprefix\url{10.1021/acs.jcim.0c01344} (\bibinfo{year}{2021}).

\bibitem{Ramsundar-et-al-2019}
\bibinfo{author}{Ramsundar, B.}, \bibinfo{author}{Eastman, P.},
  \bibinfo{author}{Walters, P.} \& \bibinfo{author}{Pande, V.}
\newblock \emph{\bibinfo{title}{Deep Learning for the Life Sciences: applying
  deep learning to genomics, microscopy, drug discovery, and more}}
  (\bibinfo{publisher}{O'Reilly Media}, \bibinfo{year}{2019}).

\bibitem{Ribeiro2016}
\bibinfo{author}{Ribeiro, M.~T.}, \bibinfo{author}{Singh, S.} \&
  \bibinfo{author}{Guestrin, C.}
\newblock \bibinfo{title}{"why should i trust you?": Explaining the predictions
  of any classifier}.
\newblock In \emph{\bibinfo{booktitle}{Proceedings of the 22nd ACM SIGKDD
  International Conference on Knowledge Discovery and Data Mining}}, KDD '16,
  \bibinfo{pages}{1135–1144}, \doiprefix\url{10.1145/2939672.2939778}
  (\bibinfo{publisher}{Association for Computing Machinery},
  \bibinfo{address}{New York, NY, USA}, \bibinfo{year}{2016}).

\bibitem{Lundberg2017}
\bibinfo{author}{Lundberg, S.~M.} \& \bibinfo{author}{Lee, S.-I.}
\newblock \bibinfo{title}{A unified approach to interpreting model
  predictions}.
\newblock In \bibinfo{editor}{Guyon, I.} \emph{et~al.} (eds.)
  \emph{\bibinfo{booktitle}{Advances in Neural Information Processing
  Systems}}, vol.~\bibinfo{volume}{30} (\bibinfo{publisher}{Curran Associates,
  Inc.}, \bibinfo{year}{2017}).

\bibitem{Rodriguez-Perez2020}
\bibinfo{author}{Rodr{\'{i}}guez-P{\'{e}}rez, R.} \& \bibinfo{author}{Bajorath,
  J.}
\newblock \bibinfo{journal}{\bibinfo{title}{{Interpretation of Compound
  Activity Predictions from Complex Machine Learning Models Using Local
  Approximations and Shapley Values}}}.
\newblock {\emph{\JournalTitle{Journal of Medicinal Chemistry}}}
  \textbf{\bibinfo{volume}{63}}, \bibinfo{pages}{8761--8777},
  \doiprefix\url{10.1021/acs.jmedchem.9b01101} (\bibinfo{year}{2020}).

\bibitem{Jiang2021}
\bibinfo{author}{Jiang, D.} \emph{et~al.}
\newblock \bibinfo{journal}{\bibinfo{title}{{Could graph neural networks learn
  better molecular representation for drug discovery? A comparison study of
  descriptor-based and graph-based models}}}.
\newblock {\emph{\JournalTitle{Journal of Cheminformatics}}}
  \textbf{\bibinfo{volume}{13}}, \bibinfo{pages}{12},
  \doiprefix\url{10.1186/s13321-020-00479-8} (\bibinfo{year}{2021}).

\bibitem{Ying2019}
\bibinfo{author}{Ying, R.}, \bibinfo{author}{Bourgeois, D.},
  \bibinfo{author}{You, J.}, \bibinfo{author}{Zitnik, M.} \&
  \bibinfo{author}{Leskovec, J.}
\newblock \bibinfo{journal}{\bibinfo{title}{{GNNExplainer: Generating
  Explanations for Graph Neural Networks}}}.
\newblock {\emph{\JournalTitle{Advances in Neural Information Processing
  Systems}}} \textbf{\bibinfo{volume}{32}} (\bibinfo{year}{2019}).
\newblock \eprint{1903.03894}.

\bibitem{Preuer2019}
\bibinfo{author}{Preuer, K.}, \bibinfo{author}{Klambauer, G.},
  \bibinfo{author}{Rippmann, F.}, \bibinfo{author}{Hochreiter, S.} \&
  \bibinfo{author}{Unterthiner, T.}
\newblock \bibinfo{title}{Interpretable deep learning in drug discovery}.
\newblock In \emph{\bibinfo{booktitle}{Explainable {AI}: Interpreting,
  Explaining and Visualizing Deep Learning}}, \bibinfo{pages}{331--345}
  (\bibinfo{publisher}{Springer}, \bibinfo{year}{2019}).

\bibitem{Kazius2005}
\bibinfo{author}{Kazius, J.}, \bibinfo{author}{McGuire, R.} \&
  \bibinfo{author}{Bursi, R.}
\newblock \bibinfo{journal}{\bibinfo{title}{{Derivation and validation of
  toxicophores for mutagenicity prediction}}}.
\newblock {\emph{\JournalTitle{Journal of Medicinal Chemistry}}}
  \textbf{\bibinfo{volume}{48}}, \bibinfo{pages}{312--320},
  \doiprefix\url{10.1021/jm040835a} (\bibinfo{year}{2005}).

\bibitem{Yang2017}
\bibinfo{author}{Yang, H.} \emph{et~al.}
\newblock \bibinfo{journal}{\bibinfo{title}{{Evaluation of Different Methods
  for Identification of Structural Alerts Using Chemical Ames Mutagenicity Data
  Set as a Benchmark}}}.
\newblock {\emph{\JournalTitle{Chemical Research in Toxicology}}}
  \textbf{\bibinfo{volume}{30}}, \bibinfo{pages}{1355--1364},
  \doiprefix\url{10.1021/acs.chemrestox.7b00083} (\bibinfo{year}{2017}).

\bibitem{Hevener2018}
\bibinfo{author}{Hevener, K.~E.}
\newblock \bibinfo{title}{{Computational toxicology methods in chemical library
  design and high-throughput screening hit validation}}.
\newblock In \emph{\bibinfo{booktitle}{Methods in Molecular Biology}}, vol.
  \bibinfo{volume}{1800}, \bibinfo{pages}{275--285},
  \doiprefix\url{10.1007/978-1-4939-7899-1_13} (\bibinfo{publisher}{Humana
  Press Inc.}, \bibinfo{year}{2018}).

\end{thebibliography}

\end{document}


\flushbottom

%
%
\thispagestyle{empty}
\renewcommand{\thefigure}{S\arabic{figure}}

{\raggedright\sffamily\bfseries\fontsize{20}{25}\selectfont{Supplementary}\par}

\section{Related Work}

\begin{figure}[!htb]
    \vspace{-0.5em}
    \centering
    \makebox[0pt]{%
    \includegraphics[scale=0.7]{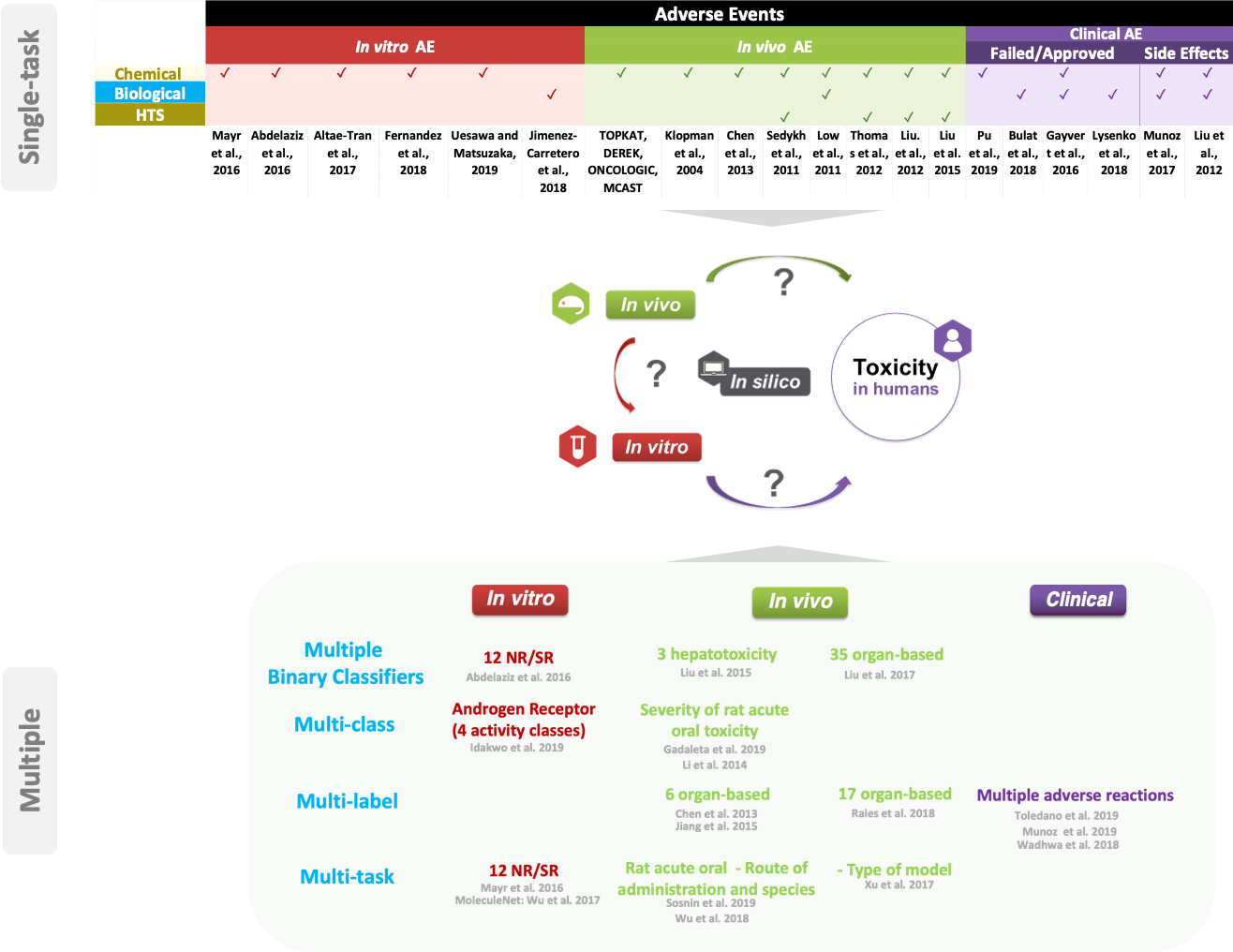}}
    \caption{\textbf{Problems in concordance and variety of input data used to predict a broad spectrum of endpoints across the \textit{in vitro}, \textit{in vivo} and clinical platforms with both single-task and multi-task models.} Multi-class models \cite{Gadaleta2019, Li2014, Idakwo2019} have predicted multiple toxic endpoints mutually exclusively per chemical. Multi-label models \cite{Chen2013,Jiang2015, Raies2018} have further generalized allowing mutually inclusive prediction of multiple endpoints for each chemical. Multi-task models \cite{Wu2018, Sosnin2019, Xu2017} have learned from multiple tasks simultaneously and predict multiple or single endpoints across different tasks. The multi-label and multi-task models applied thus far are specific to tested platform(\textit{in vivo}, \textit{in vitro} or clinically). 
    \footnotesize{\textit{References: Liu et al. 2015 \cite{Liu2015a}, Liu et al. 2017 \cite{Liua}, Gadaleta et al. 2019 \cite{Gadaleta2019}, Li et al. 2014 \cite{Li2014}, Chen et al. 2013 \cite{Chen2013}, Jiang et al. 2015 \cite{Jiang2015}, Raies et al. 2018 \cite{Raies2018}, Sosnin et al. 2019 \cite{Sosnin2019}, Xu et al. 2017 \cite{Xu2017}, Wu et al. 2018 \cite{Wu2018}, Abdelaziz et al. 2016 \cite{Abdelaziz2016}, Idakwo et al. 2019 \cite{Idakwo2019}, Mayr et al. 2016 \cite{Mayr2016}, Wu et al. 2017 \cite{Wu2017a}, Toledano et al. 2019 \cite{Perez-ParrasToledano2019}, Munoz et al. 2019 \cite{Munoz}, Wadhwa et al. 2018 \cite{Wadhwa2018a} } }
}
    \label{fig:challenges}
\end{figure}%
\FloatBarrier

\subsection{Related Work on Molecular Toxicity  Explanation} \label{explainability}

Explainable AI (XAI) in the Drug Discovery field is constantly expanding, particularly for molecular predictions \cite{Jimenez-Luna2020}. Traditionally the focus has been on interpreting QSAR (Quantitative Structure-Activity Relationship) models \cite{Polishchuk2017}, e.g., by compositional elemental/fragment or statistical analysis. Sharma et al. \cite{Sharma2017} explained differences between predicted toxic and non-toxic chemicals elementally, while Rasmussen et al. \cite{Rasmussen2022} and Polishchuk et al. \cite{Polishchuk2013} developed methods to examine fragment contributions to various molecular predictions. Sharma et al. also statistically determined significantly discriminating input features for toxicity predictions, such as solubility of chemicals \cite{Sharma2017}. 

Deep learning models for Drug Discovery applications have been explained by feature attribution, instance-based, graph convolutional based, self-explaining and uncertainty estimation methods \cite{Jimenez-Luna2020}. Here we will discuss methods used to explain molecular predictions. Feature attribution methods explain by placing relevance on input features to a given prediction \cite{Jimenez-Luna2020}. Within feature attribution methods, gradient-based and surrogate models have been used for molecular predictions. Gradient-based methods use the gradient of the model to determine feature importance, such as toxicophores being extracted from the DNN layers of DeepTox for \textit{in vitro} toxicity predictions \cite{Mayr2016}. Graph-based neural nets have also been explained by gradient-based feature attribution methods, for \textit{in vitro} toxicity predictions \cite{Pope2019, Akita2018} and for other molecular predictions \cite{Jimenez-Luna2021}. Surrogate feature attribution models create a surrogate interpretable model which approximates the original model. Ramsundar et al. \cite{Ramsundar-et-al-2019} applied the surrogate model, LIME (local interpretable model-agnostic explanations) \cite{Ribeiro2016}, to extract toxicophores correlating to \textit{in vitro} toxicity predictions. An extension to LIME, SHAP (Shapley additive
explanations) \cite{Lundberg2017}, was applied to locate relevant substructures to compound activity predictions \cite{Rodriguez-Perez2020} or to rank different molecular descriptors \cite{Jiang2021}. Graph convolutional based XAI methods explain by identifying subgraphs within a molecular graph input correlating to a prediction. GNNExplainer, a model agnostic method extracted subgraphs (or toxicophores) for a specific \textit{in vitro} toxic prediction that agreed with known mutagenic substructures \cite{Ying2019}. Using the filters within a graph convolutional model, toxicophores were also extracted to explain \textit{in vitro} toxicity predictions \cite{Preuer2019}.

\newpage
\section{Label Distribution}
\begin{figure}[!htb]
    \vspace{-0.5em}
    \centering
    \makebox[0pt]{%
    \includegraphics[scale=0.55]{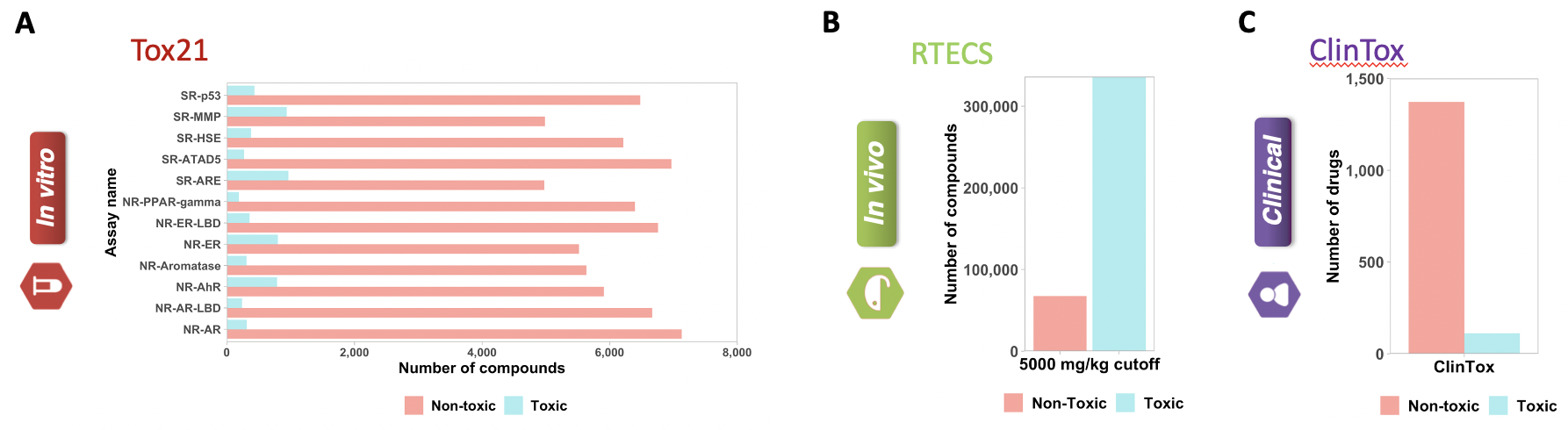}}
    \caption{\textbf{Label distribution across \textit{in vitro}, \textit{in vivo} and clinical platforms.} Toxic and non-toxic label distribution for endpoints in (A) Tox21 (\textit{in vitro}), (B) RTECS (\textit{in vivo}), and (C) ClinTox (clinical).}
\end{figure}
\FloatBarrier

\clearpage
\newpage
\section{Model Architecture}
\begin{figure}[!htb]
    \vspace{-0.5em}
    \centering
    \makebox[0pt]{%
    \includegraphics[scale=0.6]{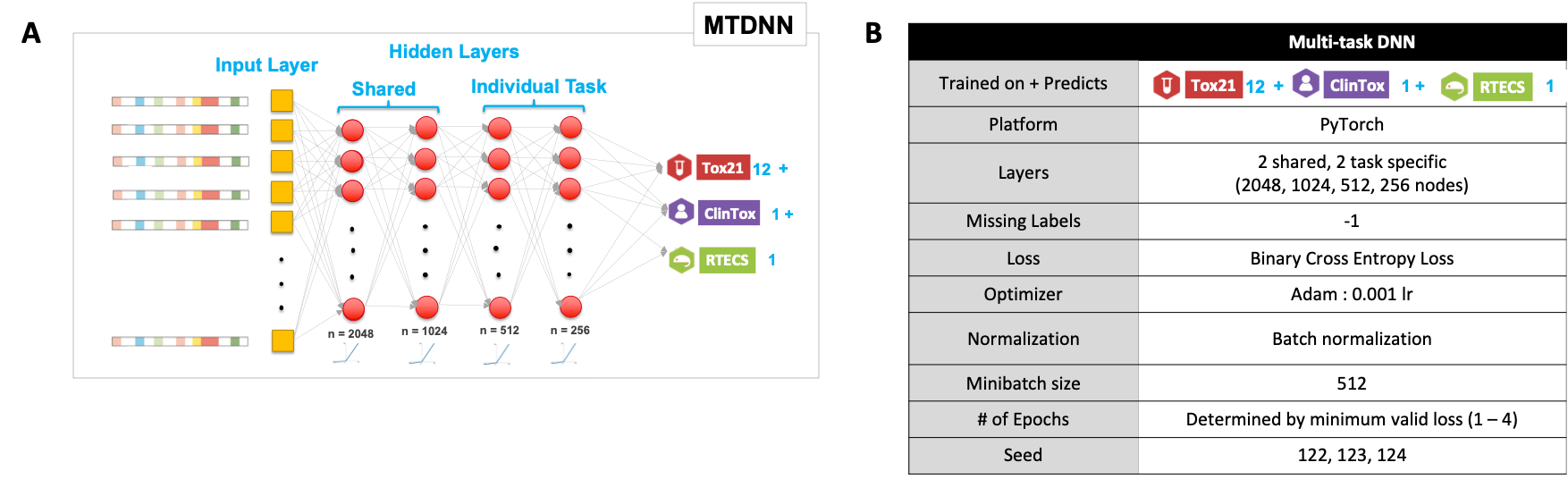}}
    \caption{\textbf{Model architecture.} (a) Multi-task Deep Neural Network (MTDNN) base architecture. Input layer of chemical fingerprints is passed through two shared layers, and two further layers for each task. Each layer undergoes batch normalization and is activated by LeakyReLU. The model combines prediction on  Tox21 (\textit{in vitro}), ClinTox (clinical) and RTECS (\textit{in vivo}) endpoints. (b) Hyperparameters used.}
    \label{fig:mtdnn_archi}
\end{figure}
\FloatBarrier

\clearpage
\newpage
\section{Full AUC-ROC comparison with baseline MoleculeNet} 
\begin{figure}[!htb]
    \vspace{-0.5em}
    \centering
    \makebox[0pt]{%
    \includegraphics[scale=0.5]{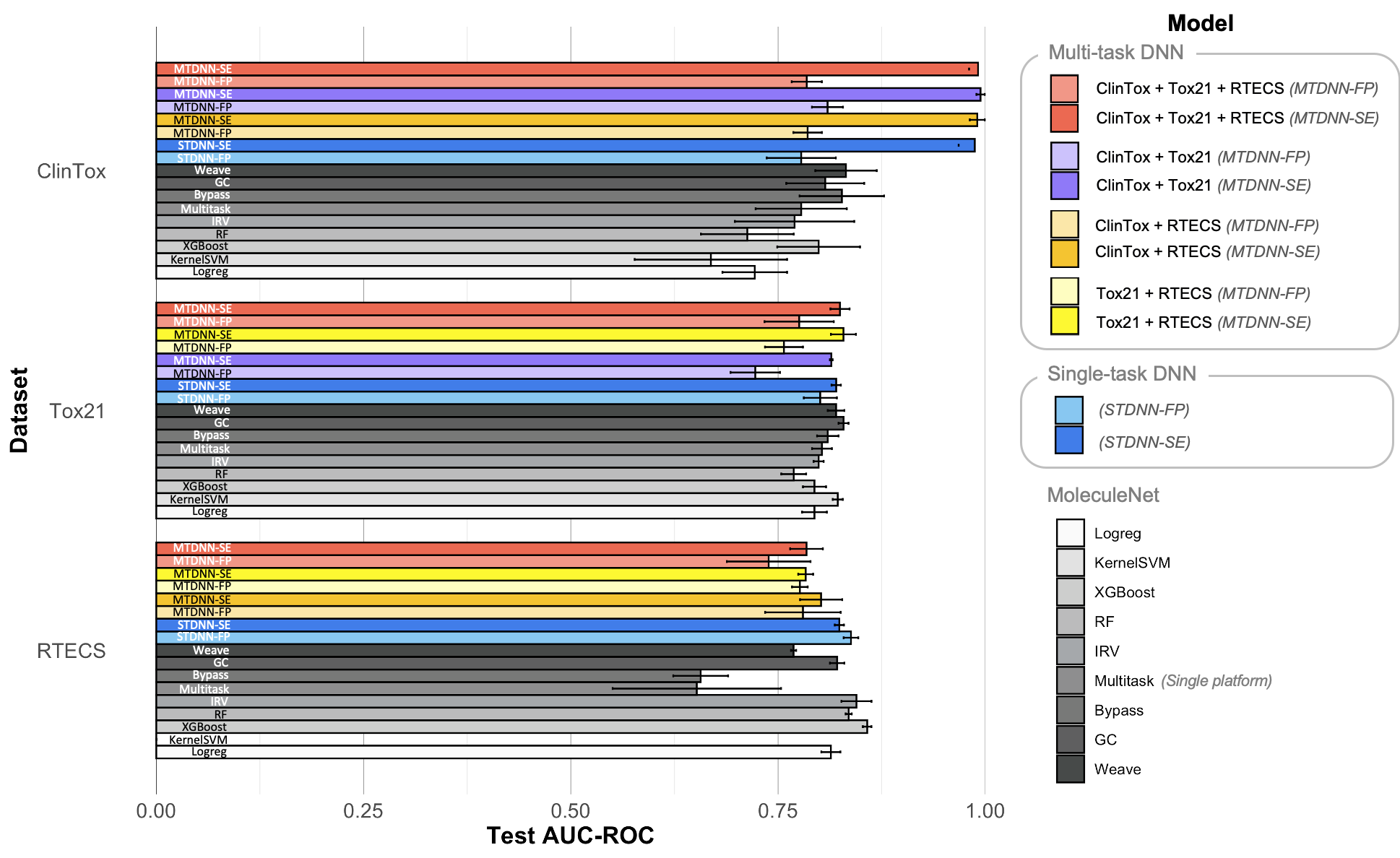}}
    \caption{ \textbf{Test AUC-ROC values for ClinTox, Tox21, and RTECS predictions, comparing multi-task models to single-task and baseline MoleculeNet models, with SMILES embeddings and Morgan fingerprints as inputs.}}
    \label{fig:auc-roc-moleculenet-embed}
\end{figure}%
\FloatBarrier

\newpage
\section{Confusion matrices comparison} 

\begin{figure}[!htb]
    \vspace{-1em}
    \centering
    \makebox[0pt]{%
    \includegraphics[scale=0.6]{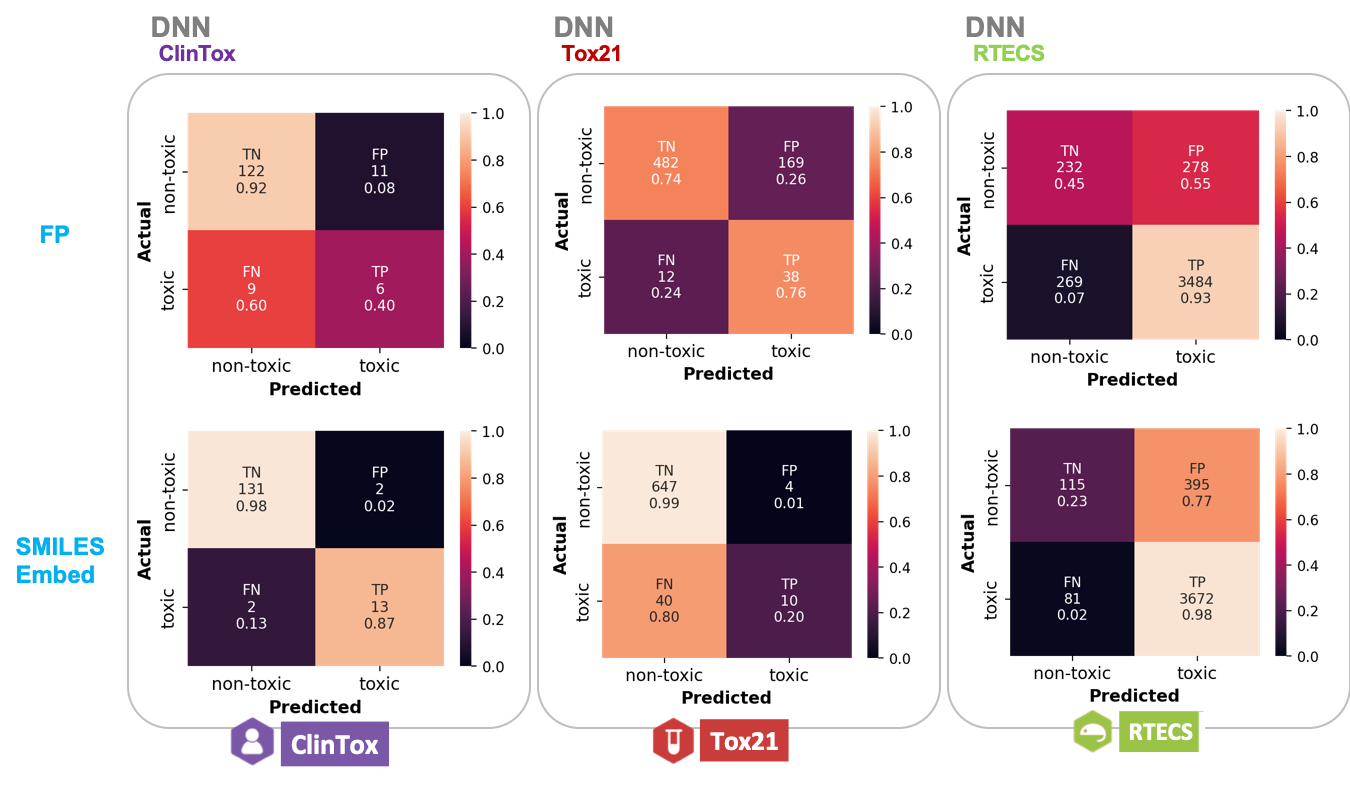}}
    \caption{\textbf{Confusion matrices for single-task DNN with Morgan fingerprints vs. SMILES embeddings as input, for ClinTox, Tox21 and RTECS tasks.} Each quadrant of the confusion matrix provides: (1) total number of chemicals, (2) normalized fraction of, true negative (TN), false positive (FP), false negative (FN) and true positive (TP) chemicals for predictions on the test dataset for separate single-task DNNs on ClinTox, Tox21 and RTECS. Given only confusion matrix results, single-task DNN performance is better for ClinTox with SMILES embedding, for Tox21 with Morgan fingerprints and for RTECS with SMILES embeddings. \textit{At seed of 122. In blue is the input molecular representation used.}}
    \label{fig:cfm-single}
\end{figure}%
\FloatBarrier

\begin{figure}[!htb]
    \vspace{-1em}
    \centering
    \makebox[0pt]{%
    \includegraphics[scale=0.6]{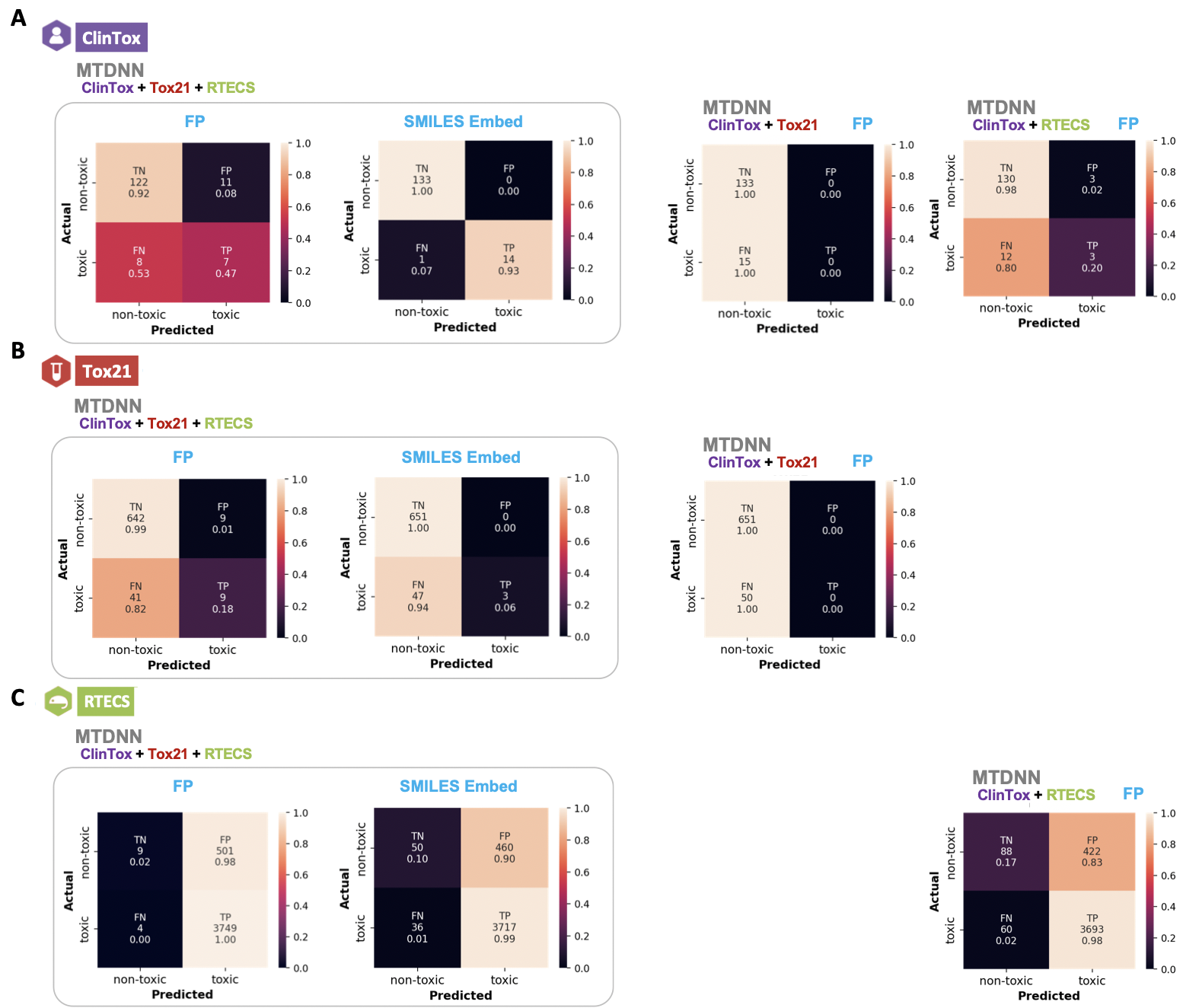}}
    \caption{\textbf{Confusion matrices for multi-task DNN with Morgan fingerprints vs. SMILES embeddings as input, for ClinTox, Tox21 and RTECS tasks.} Models are trained on the specified datasets under the ``MTDNN" label. Each quadrant of the confusion matrix provides: (1) total number of chemicals, (2) normalized fraction of, true negative (TN), false positive (FP), false negative (FN) and true positive (TP) chemicals for predictions in the specified multi-task DNN (MTDNN) tested on (a) ClinTox, (b) Tox21 and (c) RTECS. \textit{At seed of 122. In blue is the molecular representation used as input.}}
    \label{fig:cfm-multi}
\end{figure}%
\FloatBarrier

\clearpage
\newpage
\section{Full List of Top 10 Pertinent Positive and Pertinent Negative Substructures for Toxic and Non-toxic Molecules}

\begin{figure}[!htb]
    \vspace{-0.2em}
    \centering
    \makebox[0pt]{%
    \includegraphics[scale=0.75]{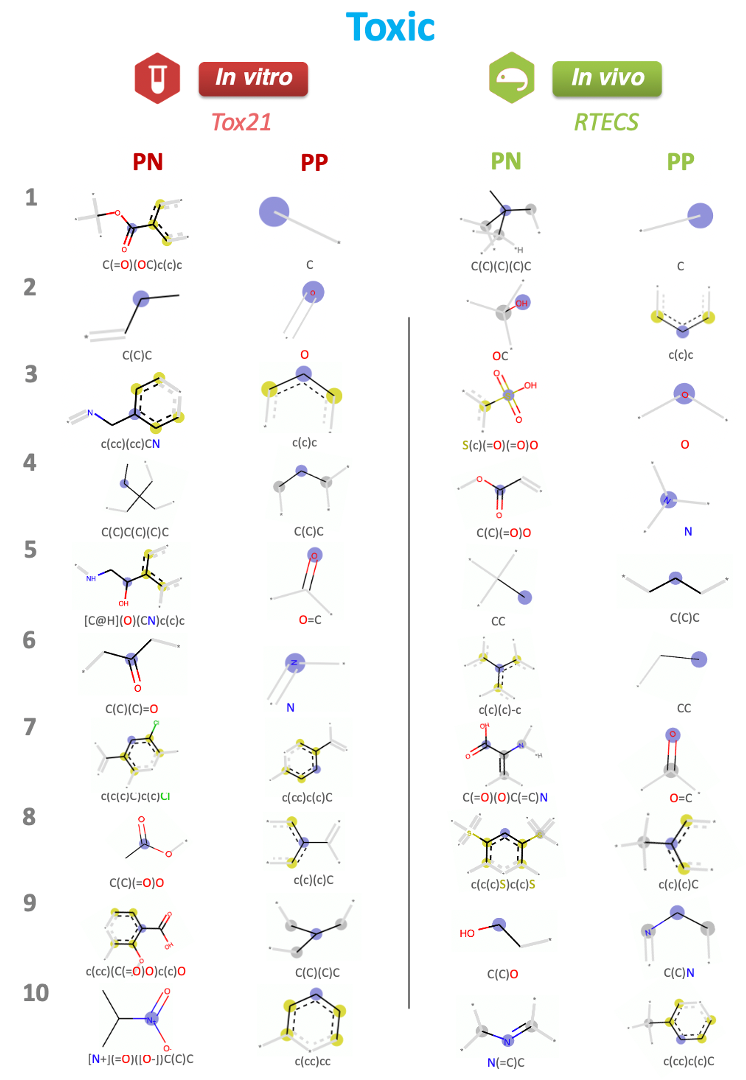}}
    \caption{\textbf{Top 10 most common PP and PN substructures of correctly predicted toxic molecules for Tox21, and RTECS endpoints.} ClinTox only had 1-2 examples of toxic molecules in the test set, and was thus excluded.}
    \label{fig:toxic}
\end{figure}%
\FloatBarrier

\begin{figure}[!htb]
    \vspace{-0.2em}
    \centering
    \makebox[0pt]{%
    \includegraphics[scale=0.75]{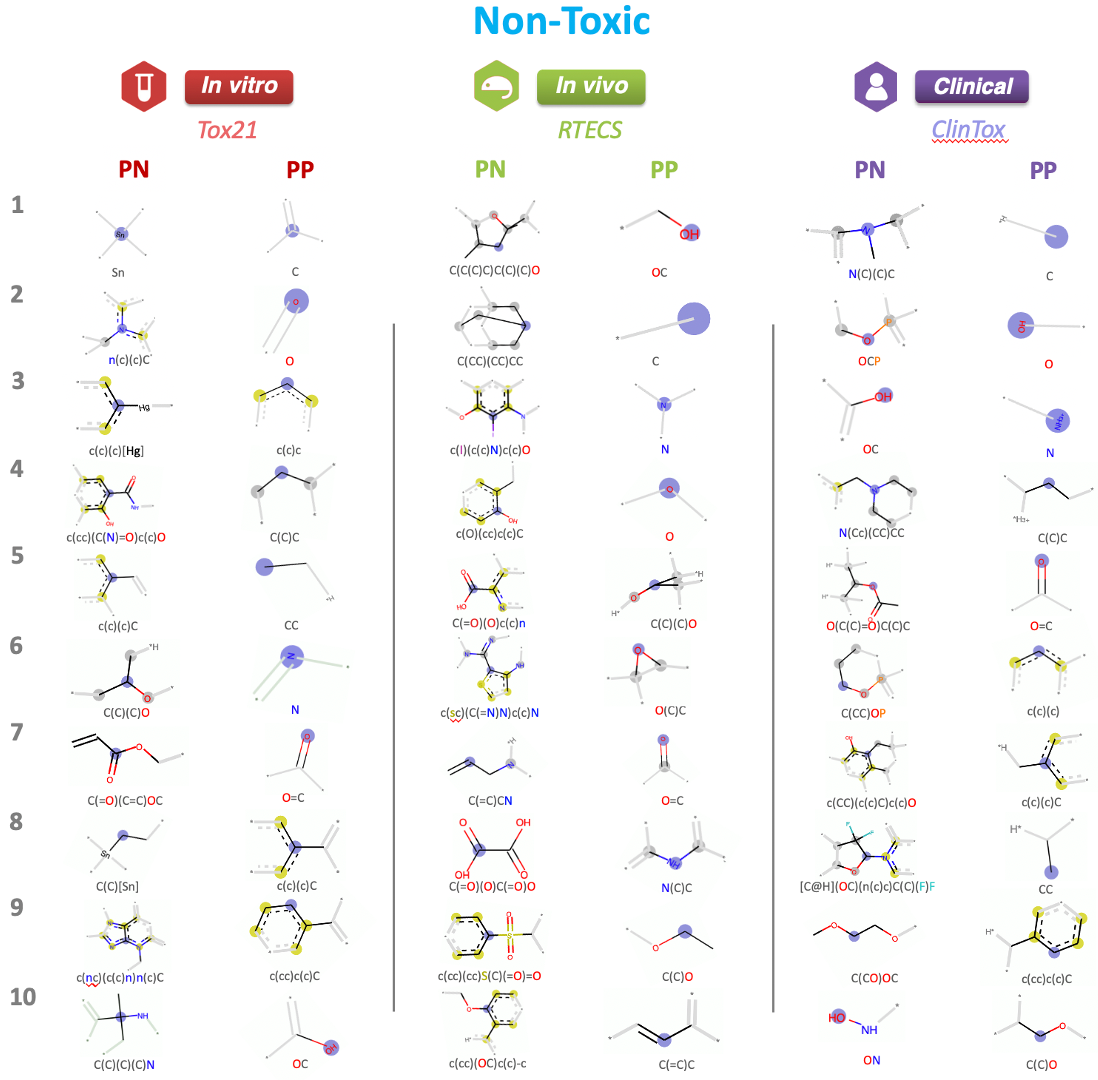}}
    \caption{\textbf{Top 10 most common PP and PN substructures of correctly predicted non-toxic molecules for Tox21, ClinTox and RTECS endpoints.}}
    \label{fig:non-toxic}
\end{figure}%
\FloatBarrier

\clearpage
\newpage
\section{Full list of Matched Toxicophores}

\begin{figure}[!htb]
    \vspace{-0.2em}
    \centering
    \makebox[0pt]{%
    \includegraphics[scale=0.6]{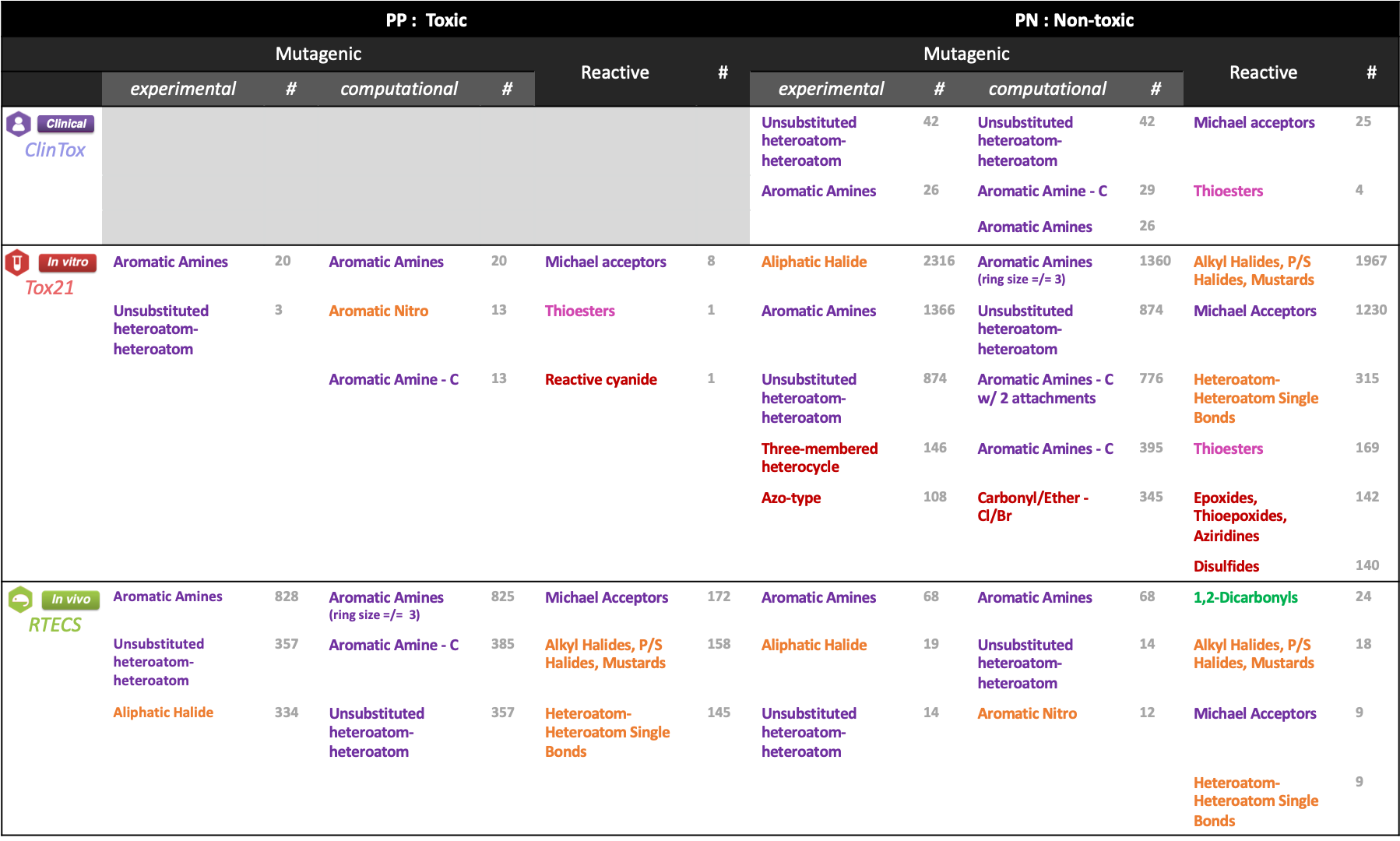}}
    \caption{\textbf{Matched Toxicophores.} Top three (ClinTox, RTECS) or top five (Tox21) matched known toxicophores to toxicophores collected from the CEM as PP of toxic molecules and PN of nontoxic molecules. For Tox21, the top five most frequent matches were examined due to the large number of matches. Three types of known toxicophores were matched: experimental \cite{Kazius2005} and computational\cite{Yang2017} mutagenic toxicophores , and reactive substructures\cite{Hevener2018} commonly used to filter molecules.}
\end{figure}%
\FloatBarrier 

\newpage
\section{Count of Matched Toxicophores Vs. Non-Toxicophores}

\begin{figure}[!htb]
    \vspace{-0.2em}
    \centering
    \makebox[0pt]{%
    \includegraphics[scale=0.4]{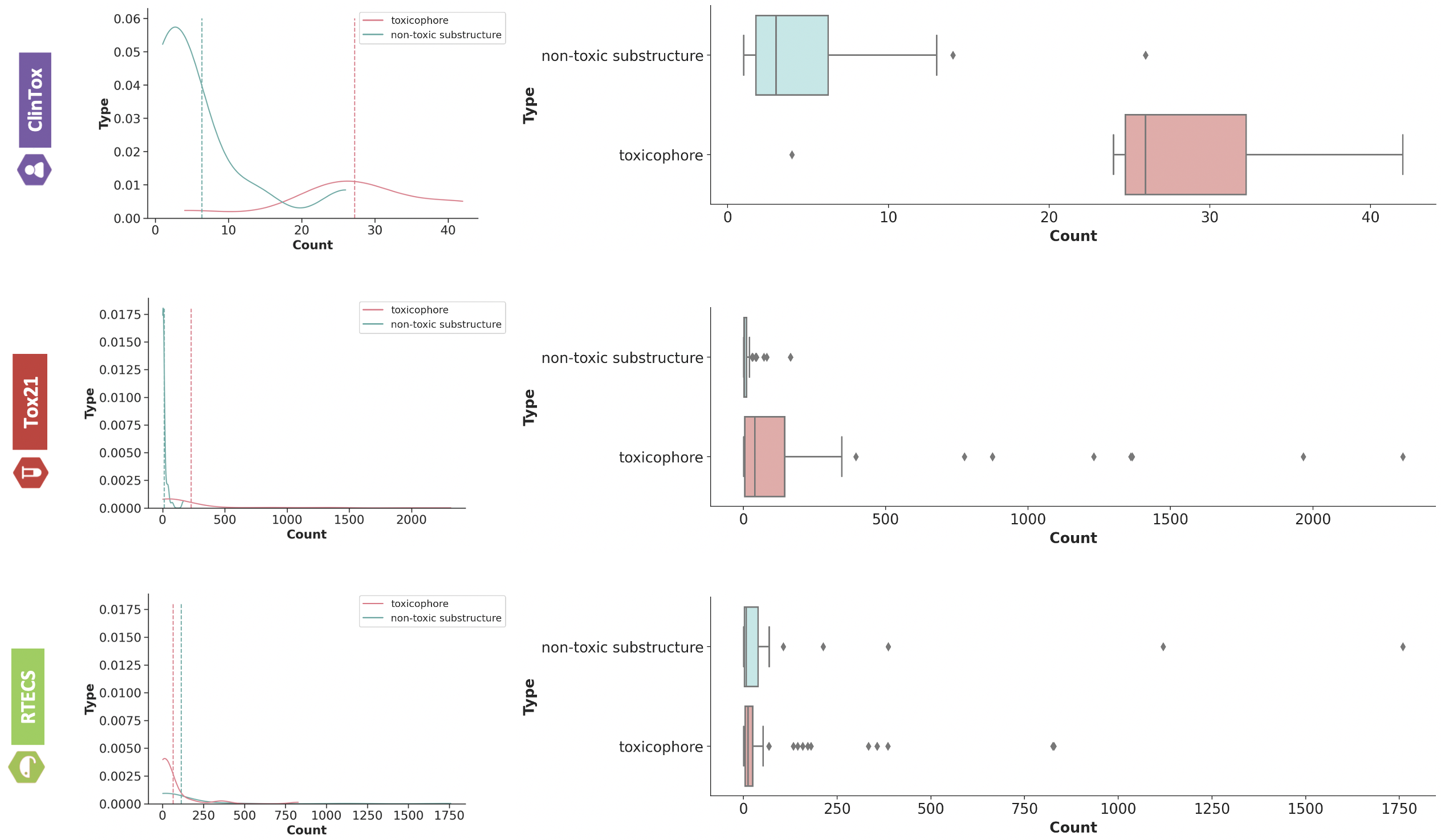}}
    \caption{\textbf{Count of toxicophores vs. non-toxic substructures obtained from the CEM that were matched to known toxicophores} All PP and PN substructures from correct predictions were matched to known toxicophores. Out of the PP and PN substructures matched to known toxicophores, the count of substructures correlating to toxic predictions (toxicophores) was compared to the count of substructures correlating to non-toxic predictions (non-toxic substurtucres). Kde-plot with mean lines, and box-plot of the counts is displayed.}
\end{figure}%
\FloatBarrier 

\newpage
\section{Cross-Confusion Matrix Comparison}

\begin{figure}[!htb]
    \vspace{-0.2em}
    \centering
    \makebox[0pt]{%
    \includegraphics[scale=0.3]{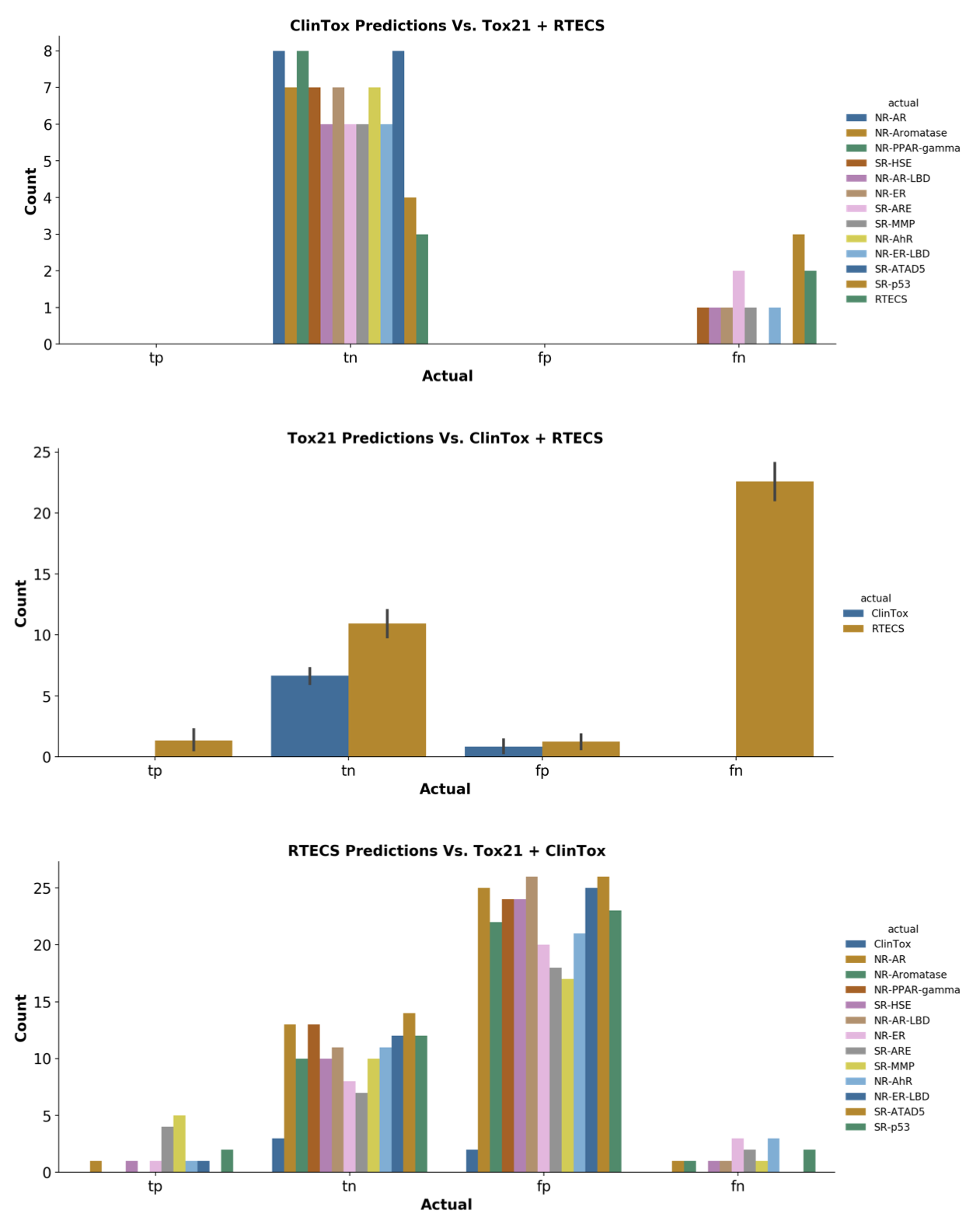}}
    \caption{\textbf{Distribution of true/false positives/negatives comparing ground truths \textit{in vitro}, \textit{in vivo} and clinically.} True positive (tp), true negative (tn), false positive (fp), and false negative (fn) comparison across predictions made for Tox21 (\textit{in vitro}), RTECS (\textit{in vivo}) and ClinTox (clinical), using ground truths across these platforms. For instance, predictions on ClinTox are compared with ground truths given by Tox21 and RTECS datasets.}
\end{figure}%
\FloatBarrier

\bibliography{biblio}